\documentclass[longauth]{aa}  

\usepackage{txfonts}
\usepackage{natbib}
\usepackage{ulem}
\usepackage{graphicx}
\usepackage[flushleft]{threeparttable}
\usepackage{multirow}
\usepackage{gensymb}
\usepackage{scrextend}   
\usepackage{hyperref}
\usepackage{xcolor}

\usepackage{hyperref}
\hypersetup{
    colorlinks=true,
    linkcolor=blue,
    filecolor=magenta,      
    urlcolor=cyan,
    citecolor=blue, 
}

\defcitealias{Franco18}{F18}  
\defcitealias{Franco2020a}{F20a} 
\defcitealias{Franco2020b}{F20b}

\begin{document} 

   \title{GOODS-ALMA: Optically dark ALMA galaxies shed light on a cluster in formation at $z$\,=\,3.5}


%

\author{L.~Zhou\inst{\ref{c1},\ref{inst1},\ref{c2}}\thanks{E-mail: \texttt{luwenjia.zhou@cea.fr}}
\and D.~Elbaz\inst{\ref{inst1}} 
\and M.~Franco\inst{\ref{inst1},\ref{inst35}} 
\and B.~Magnelli\inst{\ref{inst3}}
\and C.~Schreiber\inst{\ref{inst4}}
\and T.~Wang\inst{\ref{inst1},\ref{inst9}}
\and L.~Ciesla\inst{\ref{inst1},\ref{inst2}}
\and E.~Daddi\inst{\ref{inst1}}
\and M.~Dickinson\inst{\ref{inst5}}
\and N.~Nagar\inst{\ref{inst6}}
\and G.~Magdis\inst{\ref{inst21a},\ref{inst21b},\ref{inst21c},\ref{inst21d}}
\and D.~M.~Alexander\inst{\ref{inst8}}
\and M.~B\'ethermin\inst{\ref{inst2}}
\and R.~Demarco\inst{\ref{inst6}}
\and J.~Mullaney\inst{\ref{inst25}}
\and F.~Bournaud\inst{\ref{inst1}}
\and H.~Ferguson\inst{\ref{inst13}}
\and S.~L.~Finkelstein\inst{\ref{inst14}}
\and M.~Giavalisco\inst{\ref{inst11}}
\and H.~Inami\inst{\ref{inst15b}}
\and D.~Iono\inst{\ref{inst16},\ref{inst17}}
\and S.~Juneau\inst{\ref{inst1},\ref{inst5}}
\and G.~Lagache\inst{\ref{inst2}}
\and H.~Messias\inst{\ref{inst22},\ref{inst23}}
\and K.~Motohara\inst{\ref{inst24}}
\and K.~Okumura\inst{\ref{inst1}}
\and M.~Pannella\inst{\ref{inst10}}
\and C.~Papovich\inst{\ref{inst26},\ref{inst27}}
\and A.~Pope\inst{\ref{inst11}}
\and W.~Rujopakarn\inst{\ref{inst29},\ref{inst30},\ref{inst31}}
\and Y.~Shi\inst{\ref{c1},\ref{c2}}
\and X.~Shu\inst{\ref{inst33}}
\and J.~Silverman\inst{\ref{inst7}}}

\institute{School of Astronomy and Space Science, Nanjing University, Nanjing 210093, China \label{c1}
\and AIM, CEA, CNRS, Universit\'{e} Paris-Saclay, Universit\'{e} Paris Diderot, Sorbonne Paris Cit\'{e}, F-91191 Gif-sur-Yvette, France  \label{inst1}
\and Key Laboratory of Modern Astronomy and Astrophysics (Nanjing University), Ministry of Education, Nanjing 210093, China \label{c2}
\and Centre for Astrophysics Research, University of Hertfordshire, Hatfield, AL10 9AB, UK\label{inst35}
\and Argelander-Institut f\"{u}r Astronomie, Universit\"{a}t Bonn, Auf dem H\"{u}gel 71, D-53121 Bonn, Germany\label{inst3}
\and Department of Physics, University of Oxford, Keble Road, Oxford OX1 3RH, UK\label{inst4}
\and Institute of Astronomy, University of Tokyo, 2-21-1 Osawa, Mitaka, Tokyo 181-0015, Japan\label{inst9}
\and  Aix Marseille Univ, CNRS, CNES, LAM, Marseille, France\label{inst2}
\and  Community Science and Data Center/NSF’s NOIRLab, 950 N. Cherry Ave., Tucson, AZ 85719, USA \label{inst5}
\and Departamento de Astronom\'ia, Facultad de Ciencias F\'isicas y Matem\'aticas, Universidad de Concepci\'on, Concepci\'on, Chile \label{inst6}
\and Cosmic Dawn Center at the Niels Bohr Institute, University of Copenhagen and DTU-Space, Technical University of Denmark\label{inst21a}
\and DTU Space, National Space Institute, Technical University of Denmark, Elektrovej 327, DK-2800 Kgs. Lyngby, Denmark\label{inst21b}
\and Niels Bohr Institute, University of Copenhagen, DK-2100 Copenhagen, Denmark\label{inst21c}
\and Institute for Astronomy, Astrophysics, Space Applications and Remote Sensing, National Observatory of Athens, 15236, Athens, Greece\label{inst21d}
\and Centre for Extragalactic Astronomy, Department of Physics, Durham University, Durham DH1 3LE, UK\label{inst8}
\and Department of Physics and Astronomy, The University of Sheffield, Hounsfield Road, Sheffield S3 7RH, UK\label{inst25}
\and Space Telescope Science Institute, 3700 San Martin Drive, Baltimore, MD 21218, USA\label{inst13}
\and Department of Astronomy, The University of Texas at Austin, Austin, TX 78712, USA\label{inst14}
\and Astronomy Department, University of Massachusetts, Amherst, MA 01003, USA\label{inst11}
\and Hiroshima Astrophysical Science Center, Hiroshima University, 1-3-1 Kagamiyama, Higashi-Hiroshima, Hiroshima 739-8526, Japan\label{inst15b}
\and National Astronomical Observatory of Japan, National Institutes of Natural Sciences, 2-21-1 Osawa, Mitaka, Tokyo 181-8588, Japan\label{inst16}
\and SOKENDAI (The Graduate University for Advanced Studies), 2-21-1 Osawa, Mitaka, Tokyo 181-8588, Japan\label{inst17}
\and Universidad de Concepci\'on, Barrio Universitario, Concepci\'{o}n, Chile\label{inst22}
\and Instituto de Astrof\'{i}sica e Ci\^{e}ncias Espaciais, Observat\'{o}rio Astron\'{o}mico de Lisboa, Tapada da Ajuda, 1349-018 Lisbon, Portugal\label{inst23}
\and Institute of Astronomy, Graduate School of Science, The University of Tokyo, 2-21-1 Osawa, Mitaka, Tokyo 181-0015, Japan\label{inst24}
\and Astronomy Unit, Department of Physics, University of Trieste, via Tiepolo 11, I-34131 Trieste, Italy \label{inst10}
\and Department of Physics and Astronomy, Texas A\&M University, College Station, TX, 77843-4242, USA\label{inst26}
\and George P. and Cynthia Woods Mitchell Institute for Fundamental Physics and Astronomy, Texas A\&M University, College Station, TX, 77843-4242, USA\label{inst27}
\and Department of Physics, Faculty of Science, Chulalongkorn University, 254 Phayathai Road, Pathumwan, Bangkok 10330, Thailand\label{inst29}
\and National Astronomical Research Institute of Thailand (Public Organization), Donkaew, Maerim, Chiangmai 50180, Thailand\label{inst30}
\and Kavli Institute for the Physics and Mathematics of the Universe (WPI), The University of Tokyo Institutes for Advanced Study, The University of Tokyo, Kashiwa, Chiba 277-8583, Japan \label{inst31}
\and Department of Physics, Anhui Normal University, Wuhu, Anhui, 241000, China\label{inst33}
\and Kavli Institute for the Physics and Mathematics of the Universe (WPI), The University of Tokyo Institutes for Advanced Study, The University of Tokyo, Kashiwa, Chiba 277-8583, Japan\label{inst7}
}
\date{Received --; accepted --}

 %
  \abstract
   {Thanks to its outstanding angular resolution, the Atacama Large Millimeter/submillimeter Array (ALMA) has recently unambiguously identified a population of optically dark galaxies with redshifts greater than $z$\,=\,3, which play an important role in the cosmic star formation in massive galaxies. In this paper we study the properties of the six optically dark galaxies detected in the  69 arcmin$^2$ GOODS-ALMA 1.1mm continuum survey. While none of them are listed in the deepest $H$-band based CANDELS catalog in the GOODS-\textit{South} field down to $H$\,=\,28.16\,AB,  we were able to de-blend two of them from their bright neighbor and  measure an $H$-band flux for them. 
We present the spectroscopic scan follow-up of five of the six sources with ALMA band 4. 
All are detected in the 2\,mm continuum with signal-to-noise ratios higher than eight. 
One emission line is detected in AGS4  ($\nu_{obs}$\,=\,151.44\,GHz with a $S/N$\,=\,8.58) and AGS17 ($\nu_{obs}$\,=\,154.78\,GHz with a $S/N$\,=\,10.23), which we interpret in both cases as being due to the CO(6--5) line at $z^{\rm AGS4}_{spec}$\,=\,3.556 and $z^{\rm AGS17}_{spec}$\,=\,3.467, respectively.
These redshifts match both the probability distribution of the photometric redshifts derived from the UV to near-infrared  spectral energy distributions (SEDs) and the far-infrared SEDs for typical dust temperatures of galaxies at these redshifts.
We present evidence that nearly 70\,\% (4/6 of galaxies) of the optically dark galaxies belong to the same overdensity of galaxies at $z$\,$\sim$\,3.5. overdensity
The most massive one, AGS24 ({\it M$_{\star}$}\,=\,10$^{11.32^{+0.02}_{-0.19}}$\,M$_{\odot}$), is the most massive galaxy without an active galactic nucleus (AGN) at $z$\,>\,3 in the GOODS-ALMA field. It falls in the very center of the peak of the galaxy surface density, which suggests that the surrounding overdensity is a proto-cluster in the process of virialization and that AGS24 is the candidate progenitor of the future brightest cluster galaxy (BCG).}

\titlerunning{GOODS-ALMA: Optically dark galaxies and a cluster in formation at $z$\,=\,3.5}

   \maketitle
%
\section{Introduction}
\label{intro}
The discovery of bright submm-selected galaxies \citep[SMGs; S(850\,$\mu$m)\,>\,5mJy; e.g.,][]{Blain02} opened up a window onto the formation of the most intense episodes of star formation in massive distant galaxies. It is now widely accepted that understanding the nature of SMGs is a prerequisite to understanding the formation of local massive ellipticals and brightest cluster galaxies (BCGs) at the center of the most massive virialized structures. Defining starbursts as galaxies above the star formation main sequence (MS, star formation rate vs. stellar mass, SFR\,-\,M$_{\star}$ correlation;  \citealt{Noeske07,  Elbaz07, Elbaz11, Daddi07, Rodighiero11}), it has been found that some SMGs fall into the starburst category while some are MS galaxies (massive MS galaxies at $z$\,=\,3\,--\,5 exhibit very high star formation rates (SFRs), \citealt{Schreiber15}), some exhibit compact star formation, and others exhibit extended star formation (see e.g., \citealt{Hodge19, Rujopakarn16, Rujopakarn19, elbaz18}).

Using ALMA to follow up on 63 optically dark galaxies located in the three Southern Hemisphere CANDELS fields (totalizing 600 arcmin$^2$ in the COSMOS, EGS, and GOODS-South fields), \citet{Wang19} showed that these galaxies contribute ten times more than equivalently massive UV-bright galaxies to the cosmic SFR density at $z$\,$>$\,3.
Despite being bright -- although three to ten times fainter than the classical SMG population -- in the submillimeter (here 870\,$\mu$m), these galaxies  unseen in the deepest UV to near-infrared surveys were selected as $H$-dropouts with an Infrared Array Camera (IRAC) counterpart brighter than 24\,AB at 4.5\,$\mu$m. 
Thanks to the angular resolution of ALMA (0\farcs6 here), it is indisputable that these galaxies exhibit no counterpart from the UV to the $H$-band. While in the observations from single dish telescopes such as Submillimeter Common-User Bolometer Array (SCUBA), LABOCA LArge APEX BOlometer (LABOCA),  the IRAM\,30\,m, and Large Millimeter Telescope (LMT), several optical sources can play the role of candidate counterparts  due to the large beam size. It took 14 years \citep{Walter12} to identify the galaxy responsible for the brightest  source in the HUDF \citep[HDF850.1]{Hughes98}.
The $H$-dropout (or optically dark) galaxies followed up by ALMA experience lower, hence more normal, star formation activity than the more classical SMGs, such as those followed up by ALMA as part of the ALESS project  (see e.g., \citealt{daCunha15}, \citealt{Swinbank14}). The ALMA detected ones exhibit <SFR>\,=\,200\,M$_{\odot}$\,yr$^{-1}$ typical of MS galaxies at these redshifts although with a large scatter, whereas the undetected ones experience much lower SFRs \citep{Wang19}.


The clustering properties of these $H$-dropout galaxies -- derived from their cross-correlation with $H$-detected galaxies with similar redshifts and masses -- show that they inhabit massive dark matter halos (M$_{h}$$\sim$10$^{13\pm0.3}$$h^{-1}$M$_{\odot}$ at $z$\,=\,4) and constitute excellent candidate progenitors of the most massive galaxies presently located in the center of massive groups and galaxy clusters \citep{Wang19}.
Recent simulations strengthen this hypothesis by showing that galaxies in (proto)clusters at $z$\,>\,3 may produce as much as 50\,\% of the cosmic SFR density \citep{Chiang17}. 
Numerical simulations can be used to trace back the positions of the galaxies that will end up in a present-day galaxy cluster.  \citet{Chiang2013} followed up these galaxies in the stage  preceding virialization, which they called the proto-cluster phase. They also defined a useful reference effective radius  encompassing 65\% of the mass that will end up in the cluster.  At a redshift of $z$\,$\sim$\,3, which will be of interest in our analysis, this radius ranges from 5 comoving-Mpc (hereafter cMpc) for a M($z$\,=\,0)\,=\,3\,$\times$\,10$^{14}$\,M$_{\odot}$ bound halo to >\,8 cMpc for a Coma-like M($z$\,=\,0)\,=\,10$^{15}$\,M$_{\odot}$ halo. This is in line with the enclosed sizes, which contain 90\% of the stellar mass in proto-clusters, found in \citet{Muldrew2015}. Identifying a proto-cluster from observations is much more complex since it is difficult to determine the future of any given overdensity of galaxies. However, when studying galaxy overdensities the effective radius defined from numerical simulations may be used as a reference size for a potential proto-cluster.
Based on the standard hierarchical models of galaxy formation, numerical simulations demonstrate that massive galaxies at high redshifts are useful tracers of overdense regions since they preferentially form within the peaks in the density field  \citep{Springel05, Lucia07}. At the same time, recent ALMA surveys have revealed that  ALMA is a powerful tool for detecting  massive galaxies at high redshift \citep{Dunlop17, Hatsukade2018,  Franco18} taking advantage of its high sensitivity and high spatial-resolution. 
As a result, not only can such a population represent a unique probe of star formation at cosmic epochs when dust attenuation is largely unknown (i.e., at $z$\,$>$\,3), but it can also be seen as a potential beacon to identify clusters at the epoch of their formation. 
Studying the connection of optically dark ALMA galaxies with the formation of large-scale structures is the main goal of the present paper.

The study on how galaxy clusters form and evolve across cosmic time is critical for  testing models  of large-scale structure formation. The formation of galaxy clusters is related to the collapse of the gravitationally bound overdensities originating from the peaks in the initial density fluctuations since the Big Bang \citep{Kravtsov12}. It is also well known that galaxies evolve differently in different environments at early times: Galaxies in denser regions tend to form earlier and evolve faster, as indicated by their ages and enhanced SFR,  than galaxies in less dense regions \citep{Thomas05, Elbaz07, Gobat2008}. Therefore, by exploring the assembly of the most massive dark matter halos we  hope to better understand how  massive galaxies formed, in the framework of large-scale structure formation.

Various techniques have been developed to search for high-redshift galaxy clusters. 
However, a complete search for galaxy clusters at high redshift ($\gtrsim$2) becomes more difficult as the cluster properties evolve dramatically at an early age. Some galaxy clusters are mature systems with evidence of quiescent galaxies dominating the center \citep{Papovich12, Newman14, Willis2020}, while others still show substantial star formation activities in the center \citep{Brodwin13, Webb15, Wang16}.   As a result, infrared and submilimeter  surveys can be beneficial in revealing the overdensities \citep{Daddi09, Capak11, Miller2018, Oteo2018, Gomez-Guijarro2019, Castignani2020}, and extensive in-depth deep surveys from Atacama Large Millimeter Array (ALMA) can make significant contributions.


In this paper, we present the physical properties of six optically dark galaxies detected in the GOODS-ALMA field based on the new ALMA data obtained from the spectroscopic follow-up of five of these galaxies. 
We also discuss the connection between their optically dark nature and the dense environment they reside in.


The GOODS-ALMA survey makes use of the 69 arcmin$^2$ ALMA image covering the deepest region of the Great Observatories Origins Deep Survey South (GOODS--{\it South}) field.  In this large blind survey, massive (>10$^{10.5}$\,M$_{\odot}$) and dust-rich galaxies at high redshift are efficiently detected due to their low surface number density in the early Universe. For more details of this survey we refer the readers to (\citealt{Franco18}, hereafter \citetalias{Franco18}).

We use cosmological parameters of H$_{0}$ = 70\,km\,s$^{-1}$\,Mpc$^{-1}$, $\Omega_{M}$ = 0.3,
and $\Omega_{\Lambda}$ = 0.7. A \citet{Salpeter55} initial mass function (IMF) is
adopted to derive stellar masses and SFRs. We convert values obtained by other studies from a \citet{Chabrier03} IMF to a \citet{Salpeter55} IMF by multiplying SFR and stellar masses by the same factor, 1.74.

\section{Data and observations}
\label{sec:obs_data}
\subsection{ALMA data and observations}
We present in the following the GOODS-ALMA survey data and the ALMA spectroscopic scan data we use for the optically dark galaxies .

\subsubsection{GOODS-ALMA survey data}
\label{sec:GS}
We use data from GOODS-ALMA, a 1.1mm cosmological survey of 69 arcmin$^2$ (6\farcm9$\times$10$\arcmin$) in the GOODS--\textit{South} field (program 2015.1.00543.S, PI: D.Elbaz). GOODS-ALMA reaches an rms sensitivity of $\sigma$\,$\simeq$\,0.18\,mJy\,beam$^{-1}$ in the 0\farcs60 tapered mosaic  \citepalias{Franco18}.  The comparison between the redshift distribution of the  detections in the GOODS-ALMA  survey \citepalias{Franco18, Franco2020a} and the one in the smaller pencil beam ALMA survey of the \textit{Hubble} Ultra Deep Field (4.5 arcmin$^2$, \citealt{Dunlop17}) shows that, for the same total observing time of about 20 hours, the shallower but 15 times wider GOODS-ALMA survey enables the detection of systematically more distant galaxies  \citepalias{Franco18, Franco2020a}. This is a natural consequence of the fact that high-redshift dusty galaxies are found to be massive. The combination of the low surface density and brightness of these galaxies then favors shallower and wider surveys.

A total of 35 galaxies have been detected above 3.5$\sigma$ in GOODS-ALMA  (\citealt{Franco2020a}, hereafter \citetalias{Franco2020a}). These include 19 galaxies detected above the 4.8$\sigma$ limit from a blind detection approach  \citepalias{Franco18} and 16 galaxies within 3.5 and 4.8$\sigma$ detected using ancillary information, mainly \textit{Spitzer}-IRAC prior positions \citepalias{Franco2020a}. The median redshift and stellar mass of the S/N\,$\geq$\,4.8 sources are $z$\,=\,2.73 and {\it M$_{\star}$}\,=\,1.0\,$\times$\,10$^{11}$\,M$_{\odot}$, whereas the sources with a 4.8\,$>$\,S/N\,$\geq$\,3.5, are both slightly closer ($z$\,=\,2.40) and less massive ({\it M$_{\star}$}\,=\,7.2\,$\times$\,10$^{10}$\,M$_{\odot}$).

\subsubsection{ALMA follow-up observations of five optically dark galaxies}
\label{sec:specscan}
Out of the 35 ALMA galaxies presented in Section~\ref{sec:GS}, an ensemble of six galaxies are optically dark, or also called HST-dark since they are not detected in the optical to near-infrared HST images down to a 5$\sigma$ {\it H}-band detection limit (HST-WFC3/F160W) of $H$\,=\,28.16\,AB mag \citep{Guo13}. 
As a result, about 17\% of the sources detected above 3.5$\sigma$ in the GOODS-ALMA blind survey are optically dark (and 21\% above 4.8$\sigma$).
Out of these six sources initially classified as optically dark, we will show in Section~\ref{sec:AGS4} and \ref{sec:opticaldark}  that two are associated with an $H$-band counterpart.  They were missed from the $H$-band catalog  due to blending.

We have obtained ALMA spectroscopic follow-up data for five of these optically dark galaxies (AGS4, 11, 15, 17, 24) that we discuss in  Section\,\ref{sec:hst-dark}. The sixth one, AGS25, was identified after the ALMA call hence no follow-up data have been obtained for this one. 

We followed up these five optically dark sources with ALMA band 4 using the spectroscopic scan mode in Cycle 7 (project 2018.1.01079.S PI: M. Franco). The band 4 observations were performed by combining four tunings to cover a total frequency range of 137.2--165.8\,GHz that ensures a broad redshift coverage from $z$\,=\,2 to $z$\,=\,5, with at least one line detected among the possible CO transitions from $J$\,=\,4 to $J$\,=\,7, CI and H$_2$O. The observations were carried out with the array configuration C43-4, with a synthesized beam of $\sim$0\farcs88. Each source was observed for 13 min on-source time in each tuning, reaching a typical rms sensitivity of 0.2 mJy per 400 km\,s$^{-1}$. 
We reduced the raw data using the standard ALMA pipeline with  Common Astronomy Software Application package  \citep[CASA, ][]{McMullin2007}. Then we converted the data into \textit{uvfits} format and performed analysis with the IRAM GILDAS tool \citep{Gildas} using the same method as in \citet{Jin2019}.



\subsection{Ancillary data}
\label{sec:ancillary}
GOODS-ALMA benefits from abundant ancillary data. We summarize in the following Sections the data that we use in the present study. For more details we refer the reader to \citet[Section 2.4.]{Franco18}.
\subsubsection{Optical/near-infrared data}
We use the data from the Cosmic Assembly Near-IR Deep Extragalactic Legacy Survey (CANDELS; \citealt{Grogin11}) obtained using images from the Wide Field Camera 3 (WFC3) and the Advanced Camera for Surveys (ACS) on board the {\it Hubble Space Telescope} (HST) in the filters: $Y_{125}$, $J_{125}$, $B_{435}$, $V_{606}$, $i_{775}$, $i_{814}$ and $z_{850}$. 
We also use images from the VLT, obtained in the $U$-band with VIMOS {\citep{Nonino09}}, and in the $K_s$-band with ISAAC \citep{Retzlaff10} and HAWK-I \citep{Fontana14} and the associated CANDELS multiwavelength catalog \citep{Guo13}. 

In the near-infrared, we also use images and data from the FourStar galaxy evolution survey (ZFOURGE, PI: I. Labb\'{e}; \citealt{Straatman16}).
These include narrow-band filter images in the $J$ ($J_1$, $J_2$, $J_3$) and $H$ ($H_s$, $H_l$) bands useful for the photometric redshift determination.
The ZFOURGE team produced the deepest $K_{s}$-band detection image by combining images from the survey itself together with images from previous surveys and reaching a 5$\sigma$ limiting depth of 26.2\,$\sim$\,26.5 AB mag in the whole Chandra Deep Field South \citep[CDFS]{Giacconi02}. 

\subsubsection{Mid-infrared data}
We use \textit{Spitzer} Infrared Array Camera (IRAC) images at 3.6 and 4.5\,$\mu$m,  reaching a depth of 26.5 AB mag,  from the ultradeep mosaics of the IRAC Ultra Deep Field \citep[IUDF]{labbe15}. These were obtained by combining the observations from the IUDF (PI: Labb\'e) and IRAC Legacy over GOODS (IGOODS, PI: Oesch) programs, together with archival data from GOODS (PI: Dickinson), S-CANDELS (PI: Fazio), ERS (PI: Fazio), and UDF2 (PI: Bouwens).

\subsubsection{Radio data}
We  use radio images at 3\,GHz and 6\,GHz from the Karl G. Jansky Very Large Array (VLA; PI: W. Rujopakarn) , which cover the entire GOODS-ALMA field. The 6\,GHz map has an angular resolution of 0\farcs31\,$\times$\,0\farcs61. The exposure time is in total 177 hours reaching an  rms noise at the pointing center of 0.32\,$\mu$Jy\,beam$^{-1}$. The 3\,GHz map has an angular resolution of 1\farcs19\,$\times$\,0\farcs59 and an  average rms noise of 1.03\,$\mu$Jy\,beam$^{-1}$. 
At 3 GHz, AGS4, 17 and 24 are detected with a $S/N$ of 10.4, 9.9, and 5.6 using \texttt{PyBDSF}\footnote{\label{pybdsf} \url{https://www.astron.nl/citt/pybdsf/}}  (Rujopakarn et al. in prep.). We examined the image and noticed that the peak fluxes of AGS11, 15 and 25 are at 3\,<\,$S/N$\,<\,5.  At 6\,GHz, only AGS4 is detected with  $S/N$\,>\,5 using \texttt{PyBDSF}\footref{pybdsf}, AGS17, 24 and 25 show peak fluxes at at 3\,<\,$S/N$\,<\,5 in the image.  We list the fluxes in Table~\ref{tab:hstdark} .
\subsubsection{X-ray data}
We use the catalog of \citet{Luo17} obtained from the 7 Msec X-ray data in the CDFS field observed in three bands: 0.5-7.0\,keV, 0.5-2.0\,keV, and 2-7\,keV.  This catalog includes a classification of candidate X-ray active galactic nuclei (AGNs) that we discuss in the following of the paper.

\subsection{Origin of the redshifts and stellar masses}
In total, 11621 galaxies located in the GOODS-ALMA field have been attributed either a photometric redshift (92\,\% of the sample) or a spectroscopic redshift (894 galaxies). We discuss below the origin of these redshifts, including the newly obtained ones from the VANDELS survey that we use to improve the contrast on the $z$$\sim$3.5 structure that we will discuss in the following sections.

\subsubsection{Photometric redshifts and stellar masses}

The photometric redshifts used in the present study come from the catalog generated by the ZFOURGE team \citep{Straatman16, Forrest17}.  We use the redshift maximizing the likelihood (minimized $\chi^2$), z$_{peak}$, as derived by fitting the observed SEDs with the code \texttt{EAzY} \citep{Brammer08}.
These photometric redshifts present an excellent agreement with existing spectroscopic redshifts with $\sigma_{z}$ = 0.020 at $z$\,>\,1.5 \citep{Straatman16}, where $\sigma_{z}$ is the normalized median absolute deviation (NMAD) of $\Delta$$z$/(1\,+\,$z$), that is, 1.48\,$\times$ the median absolute deviation (MAD) of |$z_{phot}$\,-\,$z_{spec}$|/(1\,+\,$z_{spec}$).

We adopt the stellar masses of the ZFOURGE catalog. The stellar masses  were derived with \cite{Bruzual03} stellar population synthesis models \citep{Straatman16} assuming exponentially declining star formation histories (SFH) and a \citet{Calzetti00} dust attenuation law. 

\subsubsection{Spectroscopic redshifts}
The spectroscopic redshifts are both from  the literature, as compiled in the ZFOURGE catalog, and from the latest data release from the VANDELS survey (DR3\footnote{VANDELS DR3: \url{http://archive.eso.org/cms/eso-archive-news/new-data-release-of-spectra-and-catalog-from-vandels.html}; \url{http://vandels.inaf.it/db/query_catalogs.jsp}.}). VANDELS targets star-forming galaxies at 5.5\,$\geq$\,$z$\,$\geq$\,2.5 and massive passive galaxies at 2.5\,$\geq$\,$z$\,$\geq$\,1.0 in the CANDELS CDFS and UDS fields. It is an ultra-deep spectroscopic survey of high-redshift galaxies with exposure times ranging from 20 to 80 hours per source \citep{Pentericci18, McLure18}. In total, we obtain 132 new spectroscopic redshifts, $z_{spec}$, in the GOODS-ALMA field from VANDELS and 83 of them fall in the redshift range 4\,$\geq$\,$z$\,$\geq$\,3.

\subsection{Derived parameters of the optically dark galaxies}
\subsubsection{Redshifts}
\label{sec:z}
To derive the photometric redshifts, we first de-blend\footnote{\label{de-blending}By applying the de-blending method described in \citet{Schreiber18b}, also see code developed by Corentin Schreiber: \url{https://github.com/cschreib/qdeblend}.} AGS4, 15, and 24 from their neighbors in the observed optical to mid-infrared images (ancillary data in Section~\ref{sec:ancillary}), as to be discussed in detail in Section~\ref{sec:hst-dark}.  The photometric data  are then used to  measure the  redshifts by fitting the optical to mid-infrared (rest-frame UV to near-infrared) SEDs with \texttt{EAzY} \citep{Brammer08}.  The photometric redshift of AGS25 comes from its $K_s$-band counterpart in the ZFOURGE catalog, ID$_{\rm ZFOURGE}$\,=\,11353.  AGS11 and 17 are either too faint or too close to a bright emitter in the images to be de-blended properly and derive meaningful redshifts,  therefore we do not have  photometric redshifts for these two galaxies.

The spectroscopic redshifts of AGS4 and AGS17 are obtained based on the detected emission lines in the ALMA spectroscopic follow-up and their SEDs at UV to near-infrared and far-infrared (rest-frame). This will be discussed in detail in Section~\ref{sec:Lines}. 

\subsubsection{Stellar masses}
\label{sec:mstar}
The stellar masses of  AGS4, 15 and 24 are derived from the observed optical to mid-infrared SED fitted with \texttt{FAST++}\footnote{\label{FAST++} \texttt{FAST++}: \url{https://github.com/cschreib/fastpp}, a version of the code \texttt{FAST} \citet{Kriek09} fully rewritten that can handle much larger parameter grids and offers additional features.} assuming a delayed exponentially declining SFH and a \citet{Calzetti00} attenuation law. The stellar masses of AGS11 and 17 are calculated based only on the IRAC emission. The stellar mass of AGS25 comes from its $K_s$-band counterpart in the ZFOURGE catalog.
\subsubsection{Infrared luminosities}
\label{sec:lir}
 Infrared luminosities are measured by the SED fitting of the \textit{Herschel} and ALMA flux densities for AGS4 and AGS17 using  \texttt{CIGALE} \citep{Boquien19} as in \citetalias{Franco2020b}. The other four galaxies are not detected by \textit{Herschel}, then their infrared luminosity was derived by fitting the ALMA 1.1mm point with the SED templates  in \cite{Schreiber18}, as described in \citetalias{Franco2020b}.
\subsubsection{SFRs}
\label{sec:sfr}
The SFRs are measured in  \citet[][hereafter, \citetalias{Franco2020b}]{Franco2020b}, as SFR$_{\rm tot}$  = SFR$_{\rm IR}$ + SFR$_{\rm UV}$. We calculate the SFR$_{\rm IR}$ based on the correlation between infrared luminosity and SFR in \citet{Kennicutt1998} and SFR$_{\rm UV}$ contributes only 1\% to SFR in the GOODS-ALMA detections \citepalias{Franco2020b}, therefore it is negligible.  

All six galaxies are detected in the radio at 3GHz with flux densities ranging from 4\,$\mu$Jy to 40\,$\mu$Jy (Table.~\ref{tab:hstdark}). This allows us to estimate a radio SFR. With such extremely faint flux densities,  they are not expected to present a radio excess that could be attributed to an AGN, but we checked this as follows. We calculated the luminosity at 1.4GHz (restframe) from the flux at 3GHz  assuming a radio spectral index of $\alpha$\,=\,$\alpha^{\rm 3GHz}_{\rm 1.4GHz}$\,=\,$-0.8$, the typical value for galaxies at $z$\,>\,2 \citep{Delhaize17}: 
\begin{equation}
\centering
   L_{ \rm 1.4\,GHz} \,({\rm W\; Hz^{-1}}) = \frac{4\pi D^2_L}{(1+z)^{\alpha +1}} \left(\frac{1.4}{3}\right)^{\alpha}\,S_{\rm 3GHz},
	\label{eq:S1d4GHz}
\end{equation}
where {\it D}$_{L}$ is the luminosity distance to the object. The results are listed in Table~\ref{tab:hstdark}. Then we compared the ratio $L_{\rm 1.4GHz}$/SFR with the radio excess threshold as in \citet{Delvecchio2017}. All the six optically dark galaxies have $L_{\rm1.4GHz}$/SFR well below threshold, with the one of AGS24 being five time lower and the others at least ten times lower. This is fully consistent with star-forming galaxies at $z$\,$\sim$\,3.5. A radio AGN contribution cannot be ruled out, but it must be contributing for a very small fraction to the total radio emission. Nevertheless, due to vigorous star formation dominating the total radio emission at low resolution, only Very Long Baseline Interferometry (VLBI) observations could unambiguously shed light on the possible presence of radio AGN activity on circum-nuclear (<100pc) scales.
 
We also checked in the CDFS 7Ms catalog \citep{Luo17} and found that an X-ray source is located equally close to AGS4 and another galaxy, ID$_{\rm ZFOURGE}$\,=\,12333. Therefore, it is not clear whether this X-ray source is associated with our optically dark source, AGS4.
A detailed study of the AGN contamination to the SFR of AGS4 is presented in  \citetalias{Franco2020b} where it is found to be negligible.  
  \begin{figure*}[htbp]
  \centerline{
  	\includegraphics[height=2in]{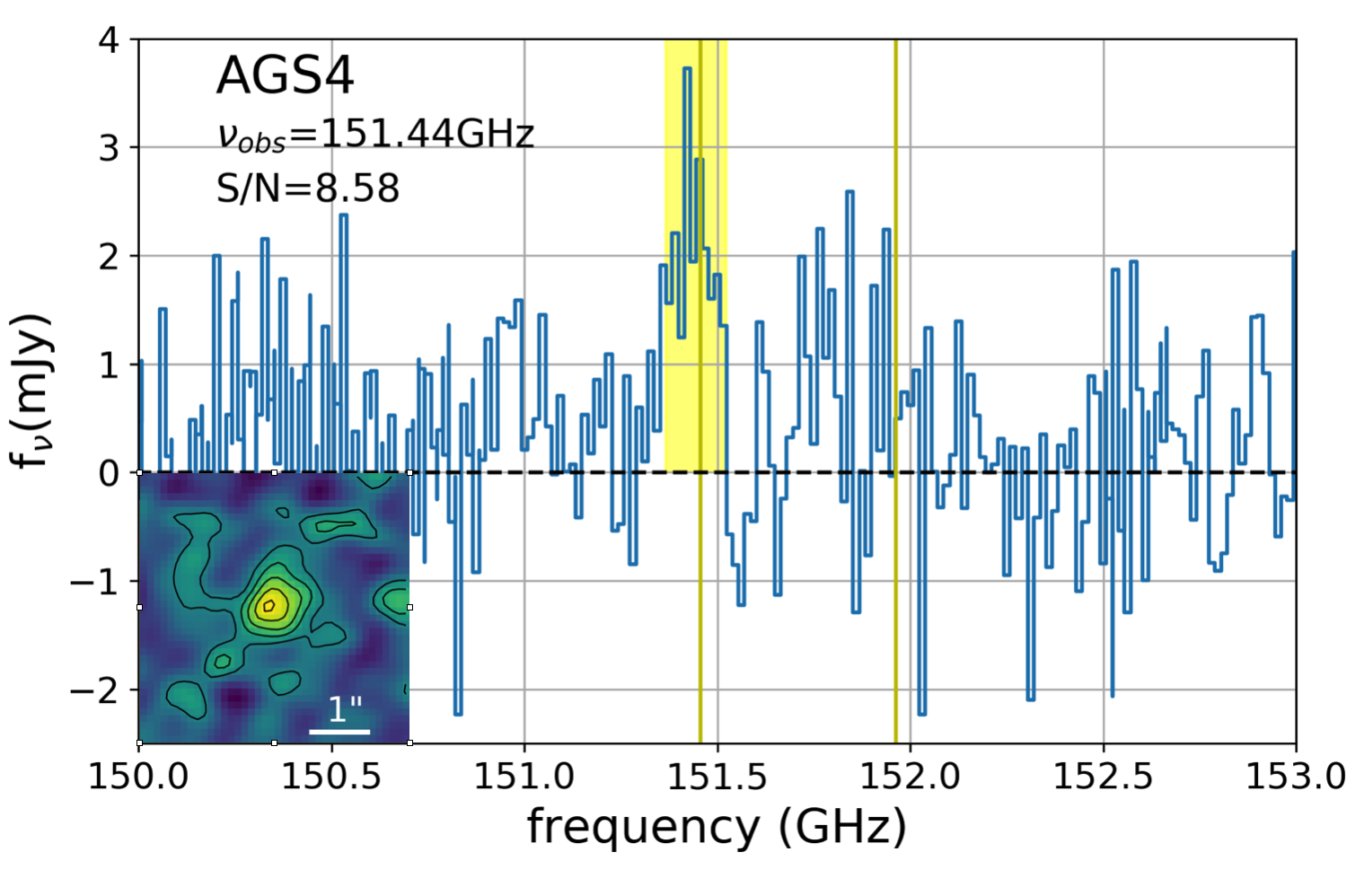}
	\includegraphics[height=2in]{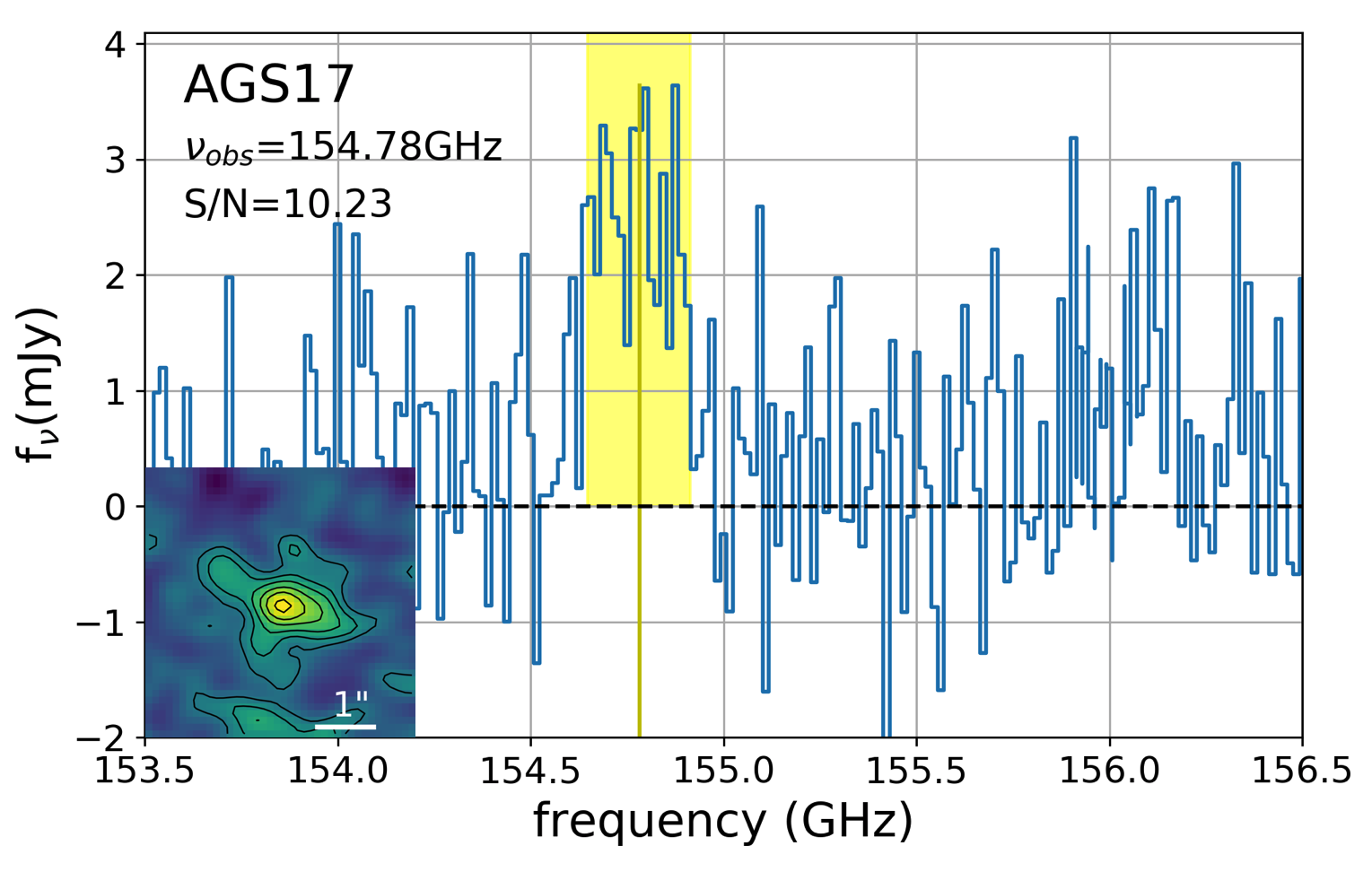}
}
\caption{
\textbf{\textit{Left:}} Part of the spectrum obtained from the ALMA band 4 observation of AGS4. The yellow shade highlights the detected emission line.  The emission line map is shown in the bottom-left corner. The contours denote 1 to 5$\sigma$, with the step of 1$\sigma$.
\textbf{\textit{Right:}} Same as on the left, but for AGS17. The contours in the map denote 1 to 6$\sigma$, with the step of 1$\sigma$. } 
 \label{fig:spec}
  \end{figure*}

\section{Results of the ALMA spectroscopic  follow-up}
\label{sec:Lines}
In this section, we present the results of the ALMA spectroscopic follow-up. As introduced in Section~\ref{sec:specscan}, five of the six optically dark galaxies are followed up, as AGS25 was identified after the ALMA call. One emission line is detected in AGS4 and AGS17, and we constrained the upper limits for AGS11, 15 and 24. 

\subsection{AGS4}
\label{sec:AGS4line}
The spectrum of AGS4 shows a clear line detection with a $S/N$\,=\,8.58 at $\nu_{obs}$\,=\,151.44\,GHz (see Fig.~\ref{fig:spec}-{\it Left}). 
One ALMA spectral line alone is not sufficient to obtain a definitive spectroscopic redshift for AGS4. However, out of the various possible lines that may be responsible for the one that is detected, only three lines can be reconciled with the 4000\,\AA\ break that is seen in the optical to near-infrared SED of AGS4 (Fig.~\ref{fig:sed}-{\it Top-left}, see more discussion on the SED in Section~\ref{sec:AGS4}), namely the CO(6-5), CO(7-6) and H$_2$O(2$_{11}$-2$_{02}$) lines.

\begin{figure}[htbp]
\centering
	\includegraphics[height=2.4in]{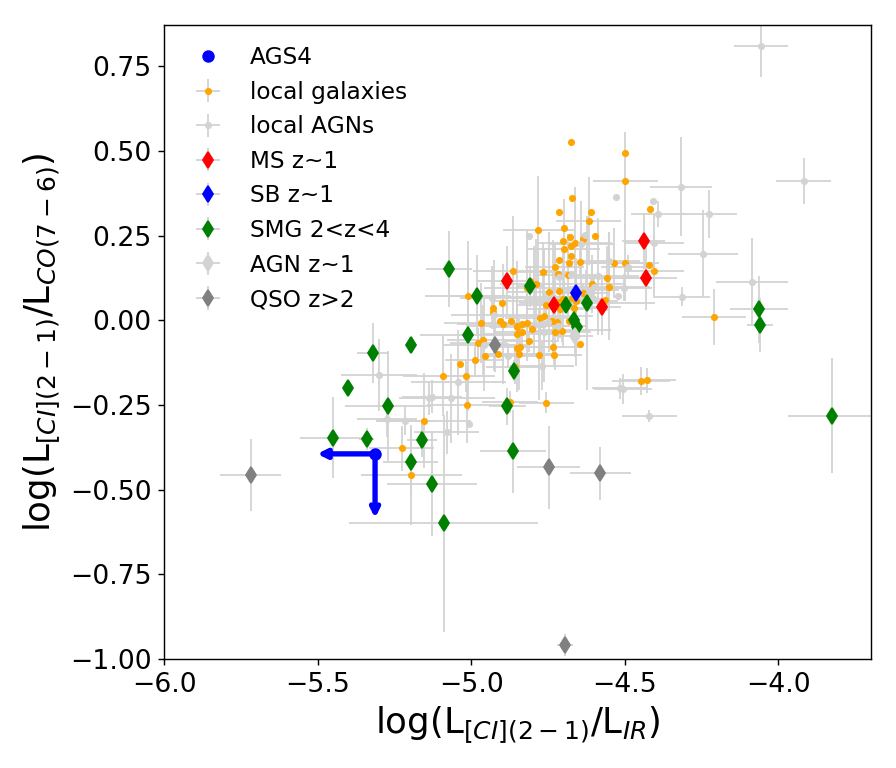} 
    \caption{ Upper limit of the [CI](2-1)/CO(7-6) flux ratio of AGS4 (blue arrows), compared with existing observations of local and high-redshift MS galaxies, starburst galaxies and AGNs (data compiled in \citealt{Valentino2018, Valentino2020}). If the detected emission line is indeed CO(7-6), then the flux ratio would fall at the lower end of the trend shown by the existing observations.}
\label{fig:fluxratio}
\end{figure}

Another reason to favor these lines comes from the far-infrared SED (Fig.22 in \citetalias{Franco18}). 
Being one of the brightest GOODS-ALMA  sources, AGS4 is detected in five \textit{Herschel} bands (at 100, 160, 250, 350, and 500\,$\mu$m), which together with the ALMA measurements, allowed \citetalias{Franco18} to obtain a robust identification of the peak of the far-infrared SED, which falls close to 350\,$\mu$m. Redshifts below $z$\,=\,2 (such as the one obtained for the 4-3 CO transition) would lead to dust temperatures (peak around 120\,$\mu$m) much colder than typical $z$$\sim$2 galaxies (see \citealt{Schreiber18}). 

We exclude H$_2$O(2$_{11}$-2$_{02}$). It is typically less bright or as bright \citep[flux ratio from 0.4 to 1.1]{Yang16} as the neighboring high-J CO transition lines (7--6 or 6--5). However, we do not find any evidence for a second line in the ALMA spectrum that can be comparable to the 8.7$\sigma$ detection.

The CO(7-6) line with a rest-frame frequency of $\nu_{\rm RF}$\,=\,806.7\,GHz provides a spectroscopic redshift of $z_{spec}^{AGS4}$\,=\,4.326, which agrees with one of the two peaks of the probability distribution function (PDF) of the optical photometric redshift. 
We note that the [CI](2-1) ($\nu_{\rm rf}$\,=\,809.3 GHz) line falls very close to the CO(7-6) line and is not detected here  with a 3.5$\sigma$ limit of 0.7\,mJy, which corresponds to a [CI](2-1)/CO(7-6) ratio of 0.40. This flux ratio falls at the lower end of the trend defined by the MS galaxies,  starburst galaxies and AGNs in the local universe and at high-$z$   \citep[data compiled and described in][]{Valentino2018,Valentino2020}, as shown in Fig.~\ref{fig:fluxratio}. 
Chance is small to have AGS4 to be an extreme case of the [CI](2-1)/CO(7-6) flux ratio. 
The CO(6-5) line with a rest-frame frequency of $\nu_{\rm RF}$\,=\,806.7\,GHz  gives a $z_{spec}^{AGS4}$\,=\,3.566, which agrees with the 4000\,\AA \  discontinuity in   the  optical SED. Therefore, we adopt the redshift to be  $z_{spec}^{AGS4}$\,=\,3.556 assuming the line to be CO(6-5).

\begin{figure*}[htbp]
\centering{
	\includegraphics[height=2.in]{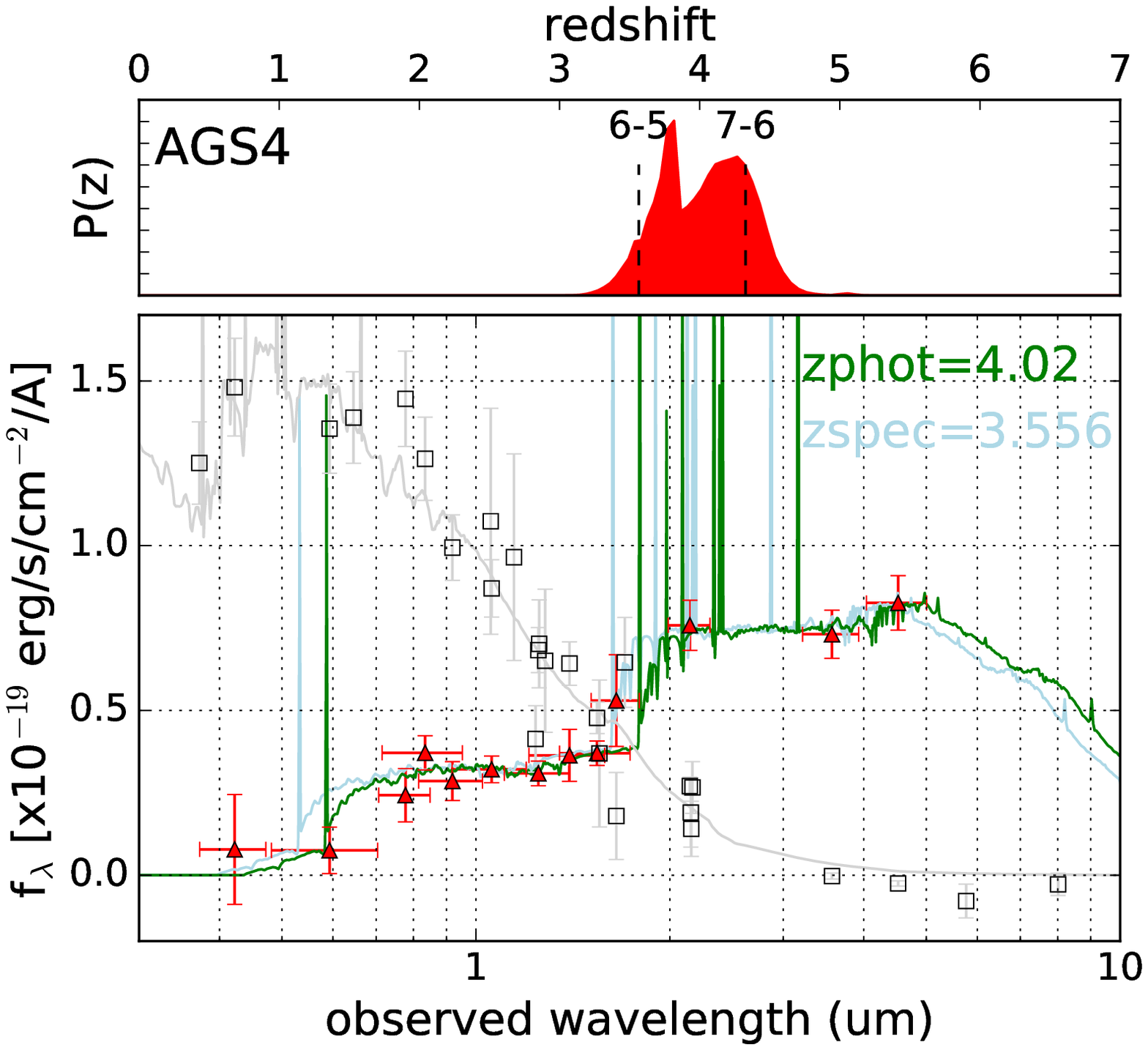}    	
	\includegraphics[height=2.in]{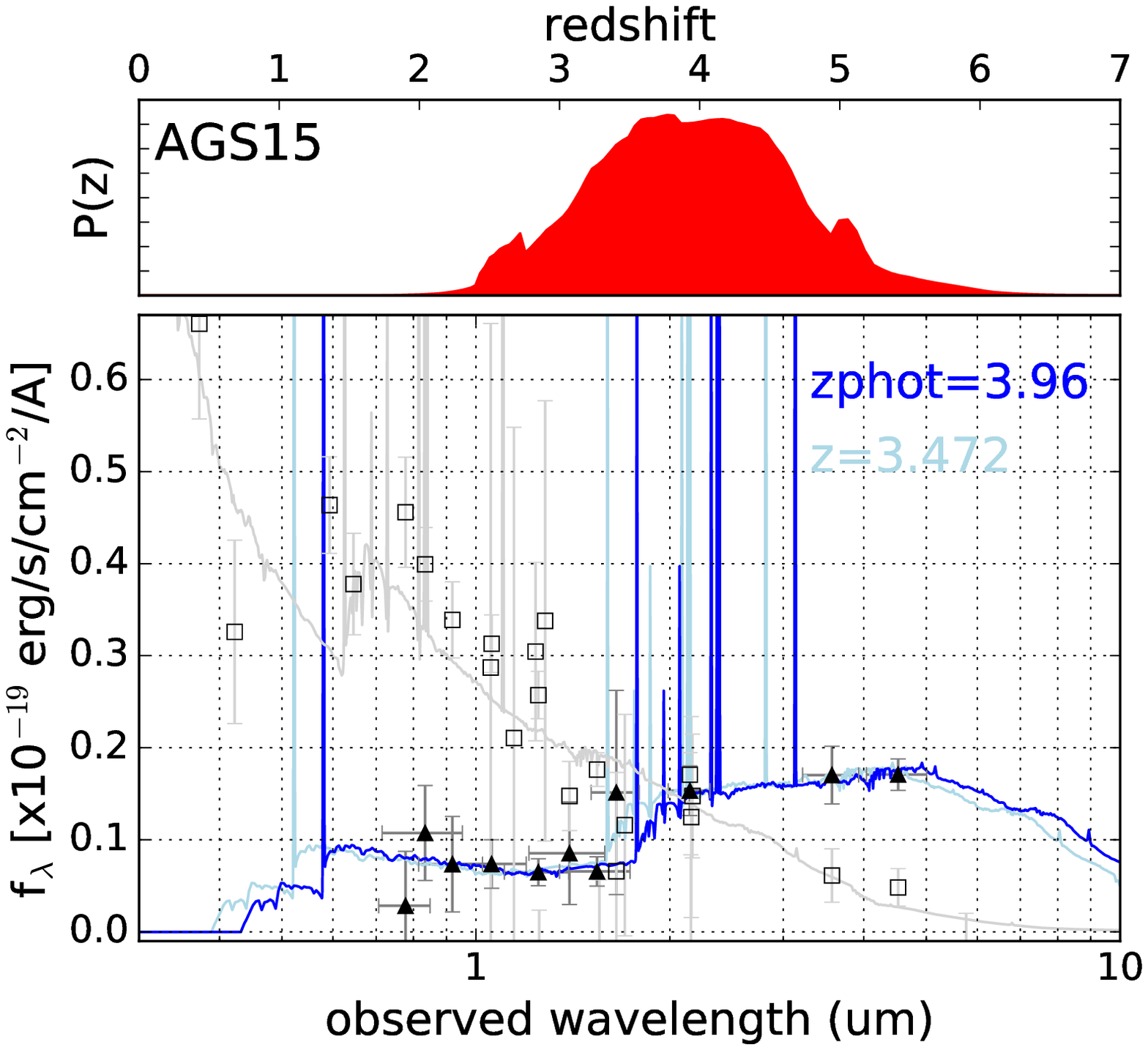}\\
	\includegraphics[height=2.in]{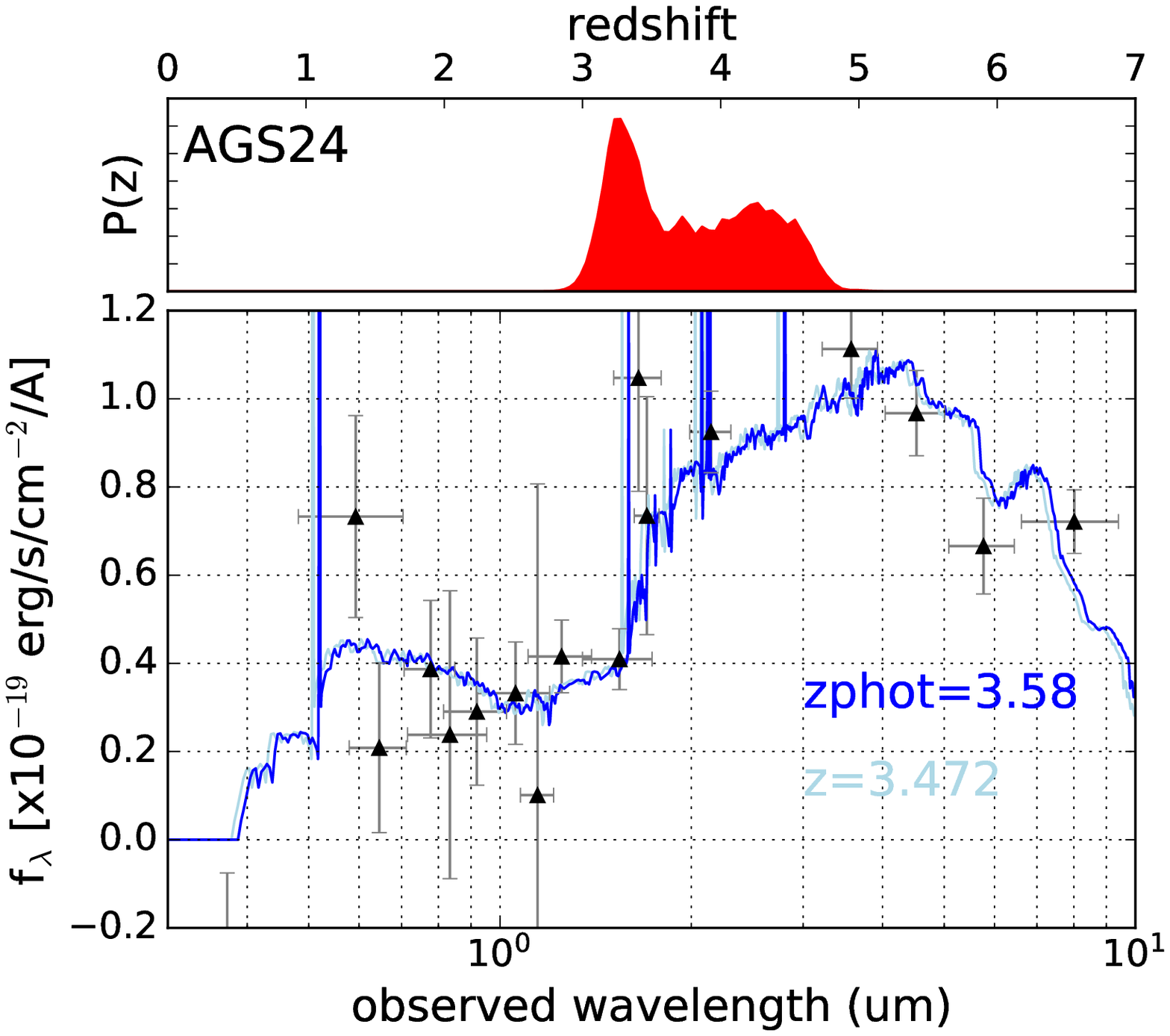}
	\includegraphics[height=2.in]{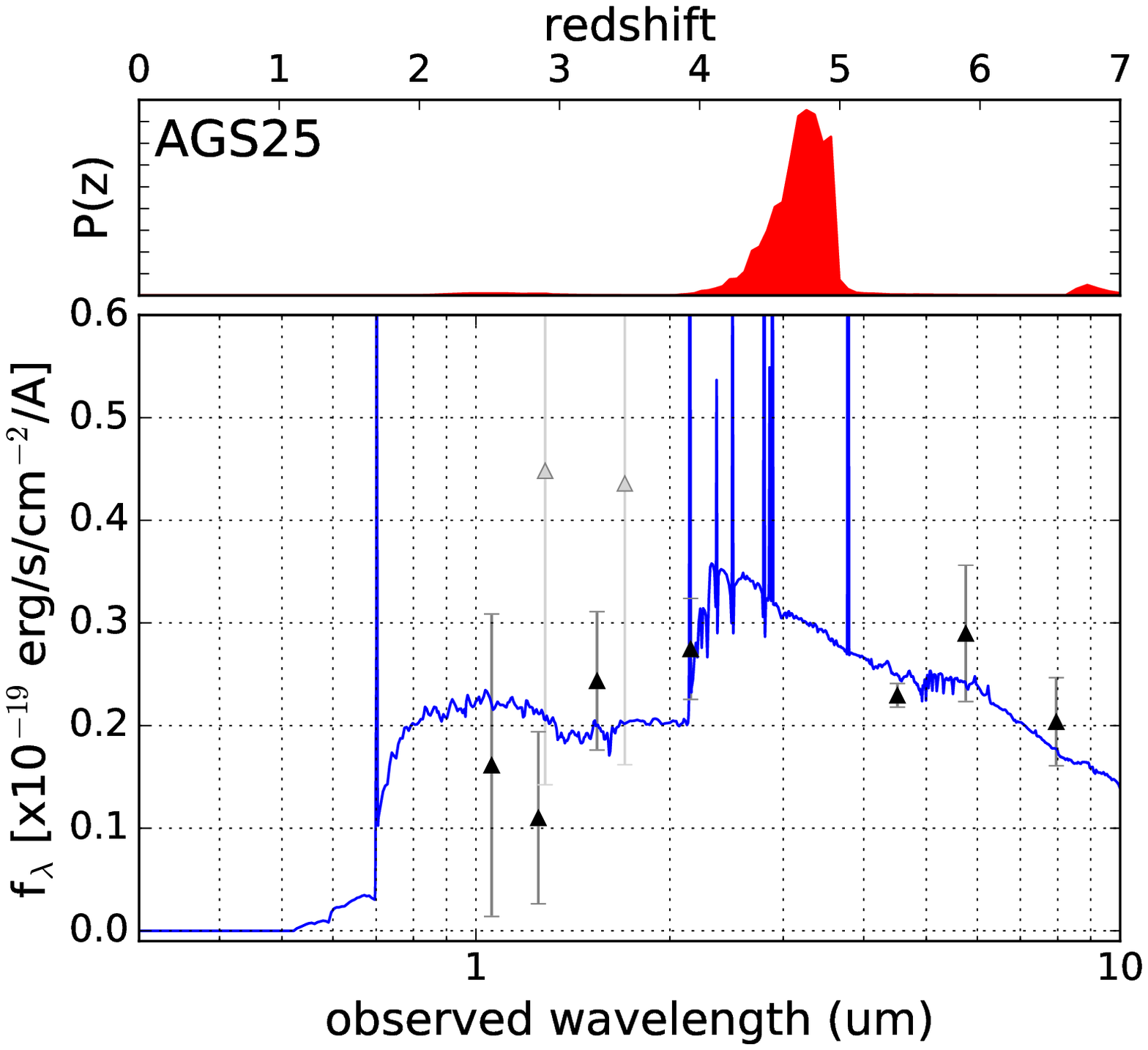}
}
   \caption{\textbf{\textit{Top-left:}} SED of AGS4.
 Top: Photometric redshift PDF derived by \texttt{EAzY} \citep{Brammer08}. 
 Bottom: The SED fittings. The green curve shows the best-fit of AGS4 at $z_{phot}$\,=\,4.02 using \texttt{EAzY}, and we shift the SED to the spectroscopic redshift ($z_{spec}$\,=\,3.556) of AGS4 in light blue for comparison.   The gray curve shows the SED of the optically bright neighbor, ID$_{\rm ZFOURGE}$\,=\,12333.
 \textbf{\textit{Top-right:}}  SED of AGS15. The blue curve shows the best-fit at $z_{phot}$\,=\,3.96 derived by  \texttt{EAzY} and we shift the SED to  the redshift peak  of the overdensity, at $z$\,=\,3.472, in light blue for comparison. The gray curve shows the SED of the optically bright neighbor, ID$_{\rm CANDELS}$\,=\,3818.
\textbf{\textit{Bottom-left:}}  SED  of AGS24. Same as AGS15. \textbf{\textit{Bottom-right:}}  SED of AGS25 at $z_{phot}$\,=\,4.70, fitted by \texttt{EAzY} with the photometry retrieved from the ZFOURGE catalog. 
}
\label{fig:sed}
\end{figure*}

\subsection{AGS17}
\label{sec:AG17line}

Our spectroscopic scan follow-up  of AGS17 shows an emission line at 154.78\,GHz with a $S/N$\,=\,10.23.
The far-infrared SED in this region  peaks at around 400\,$\mu$m (\citetalias{Franco18}, Fig.22).
Taking this far-infrared SED into account,  the detected line is  likely to be CO(5-4) or CO(6-5) or CO(7-6) since the higher-$J$ or lower-$J$ transitions give redshifts leading to unrealistic dust temperatures with dust emission peaking at $\lambda$\,<\,67\,$\mu$m or at $\lambda$\,>\,134\,$\mu$m, restframe. We checked the three $L_{\rm IR}$--$L_{\rm CO}$  relation for the three transitions. We found that the ALMA flux of AGS17 and the corresponding $L_{\rm IR}$ for the three different redshifts, which are $z$\,=\,2.723, 3.467, 4.212, will bring the galaxy to the positions that are consistent with the $L_{\rm IR}$ -- $L_{\rm CO}$ relations of the three transitions \citep{Liu2015}. Therefore, we cannot disentangle these possibilities based on the single line.

However, if we assume the line to correspond to the CO(7-6) transition ($\nu_{\rm rf}$\,=\, 806.7\,GHz), a second line, [CI] ($\nu_{\rm rf}$\,=\, 809.3 GHz), is supposed to be detected at similar significance   \citep{Valentino2020} and it is not the case for AGS17.  
We are then left with two possibilities: a spectroscopic redshift of  $z^{\rm AGS17}_{spec}$\,=\,2.723 in the case of the CO(5-4) transition and the other of  $z^{\rm AGS17}_{spec}$\,=\,3.467 for the CO(6-5) transition. The second redshift of  $z^{\rm AGS17}_{spec}$\,=\,3.467 turns out to fall within a redshift peak in the redshift distribution of the galaxies in the field, which also exhibits a concentration of galaxies right around AGS17, as we will discuss in more details in Section~\ref{sec:physics}. For that reason, despite the fact that we cannot strictly rule out the possibility that we are here observing the CO(5-4) transition, we decided to favor the CO(6-5) transition from a purely probabilistic point of view. A mild confirmation of this choice comes from the dust temperature that we obtain when assuming one or the other transition and redshift.
If  the transition is CO(5-4),  the spectroscopic redshift $z^{\rm AGS17}_{spec}$\,=\,2.723 implies a dust temperature $T_{\rm dust}$\,$\sim$\,30\,K, using a simple conversion of the peak of the SED at around 400\,$\mu$m to the dust temperature. This is lower than the typical temperature $T_{\rm dust}$\,$\sim$\,38\,K for the NS galaxies at this redshift \citep{Schreiber18}. Instead, if we assume the transition to be CO(6-5), the  $z^{\rm AGS17}_{spec}$\,=\,3.467  indicates a dust temperature $T_{\rm dust}$\,$\sim$\,37\,K that is close to the typical dust temperature $T_{\rm dust}$\,$\sim$\,40\,K  at $z$\,=\,3.5  \citep{Schreiber18} and is consistent with the mean dust temperature, $T_{\rm dust}$\,=\,37\,K, of the 39 optically dark galaxies in \citet{Wang19}. 
We note that we also tried to obtain an independent confirmation of the redshift of AGS17 from its UV to near-IR SED as we did for AGS4. However, as we will discuss in Section~\ref{sec:opticaldark} and as shown in Fig.~\ref{fig:AGS1117}-\textit{Bottom}, the optical emission of AGS17 suffers from a strong blending with bright neighbors that prevents us from extracting a meaningful SED that we could use to obtain a robust, or even tentative, model fit and photometric redshift derivation.

\subsection{Upper limits of AGS11, AGS15, and AGS24}
\label{sec:lines}
Our spectroscopic scan follow-up of AGS11, AGS15, and AGS24 does not show emission line detections higher than 5$\sigma$.  Based on their assumed redshifts, as will be discussed in Section~\ref{sec:physics},
the ALMA spectroscopic scans  encompass the position of the CO (6-5) emission line in the three galaxies.  At the assumed redshifts, the infrared luminosities can  be converted to  CO(6-5) luminosities using the relation defined by \citet{Liu2015} and then to  CO(6-5) fluxes accordingly.
We find that the non-detection of the CO(6-5) line agrees with the dispersion of the $L_{\rm IR}$-$L^{\prime}_{\rm CO\,6-5}$ relation with values below 3$\sigma$ for the three galaxies (2.8$\sigma$, 2.3$\sigma$, 0.7$\sigma$ for AGS11, AGS15, and AGS24, respectively).


%
%

\section{GOODS-ALMA optically dark galaxies}
\label{sec:hst-dark}

As described in Section~\ref{sec:GS}, six out of the 35 GOODS-ALMA detections do not have counterparts in the CANDELS catalog \citep{Guo13}.  
Despite their non detection in the $H$-band images, we are able to  measure the residual emission in the optical to near-infrared images after subtracting the contribution of neighboring galaxies and to use the resulting measurements to obtain photometric redshift determinations as discussed in the next sections. 
After a careful analysis of the $H$-band images of these galaxies, we realized that in two cases, AGS4 and AGS15, there was a clear association in the $H$-band image that was matching the ALMA contours. In both cases, the $H$-band detection was close to a bright $H$-band neighbor and both objects were interpreted in the CANDELS catalog as a single object, hence neglecting the ALMA source. If we were to extrapolate our small number statistics to the population of optically dark galaxies at these depths, this would imply that about 35\,\% of the optically dark sources are mistaken as such due to confusion. We note, however, that these galaxies are detected here because of the very deep $H$-band image since their AB magnitudes are $H$\,=\,25.23 and 27.11\,AB for AGS4, and AGS15, respectively.

All four of the optically dark galaxies detected from a blind detection above 4.8$\sigma$ are detected with IRAC and listed in the catalog of \citet{Ashby15}. These galaxies are AGS4, 11, 15 and 17. The remaining two optically dark galaxies, AGS24 and AGS25, were detected in the ALMA image using IRAC priors in \citetalias{Franco2020a}. However, they were not listed in the catalog of \cite{Ashby15}. In both cases, \citetalias{Franco2020a} (see also Section \ref{sec:massivegal}) noticed that despite being well above the IRAC detection limit, these two galaxies were missed because of the presence of a bright neighbor due to confusion (see Fig.~\ref{fig:img_AGS24}). After carefully subtracting the surrounding neighbors, the flux densities measured at 3.5\,$\mu$m for both galaxies are F$_{3.6\mu m}$(AGS24)\,=\,4.7\,$\pm$\,0.5\,$\mu$Jy and F$_{3.6\mu m}$(AGS25)\,=\,1.5\,$\pm$\,0.1\,$\mu$Jy.

Five of the GOODS-ALMA optically dark galaxies (AGS4, 11, 15, 17 and 24) were followed up with ALMA band 4 (see Section~\ref{sec:specscan}) and all are confirmed with a $S/N$ of 14.7, 16.8, 15.3, 15.9 and 8.7 in the 2\,mm continuum.  AGS25 was not included in the ALMA proposal because the list of sources below the 4.8$\sigma$ limit that includes it was still under construction at the time. All the six optically dark galaxies are also detected above 4.5$\sigma$ at 870\,$\mu$m continuum \citep{Cowie2018}. Hence there is no doubt that they are all real.

Another confirmation of the robustness of three of the optically dark galaxies -- AGS4, 17 and 24--  comes from their detection in the radio with the VLA at 3 GHz with a $S/N$ of 10.4, 9.9, and 5.6,  respectively. AGS11, 15 and 25 are not detected with $S/N$ higher than 5 (Rujopakarn et al. in prep.). The first two of the detected galaxies, AGS4 and AGS17, are the only optically dark galaxies in the sample that are also detected with \textit{Herschel} and their far-infrared SED is well-measured with no less than five \textit{Herschel} data points from 100 to 500\,$\mu$m in the observed frame. We discuss in the following two sections the properties of AGS4 and AGS25.

\subsection{AGS4, an extremely massive galaxy at $z$=3.556 and a case of blending in the Hubble $H$-band image} 
\label{sec:AGS4}
\begin{figure*}[htbp]
\centering
	\includegraphics[height=2.3in]{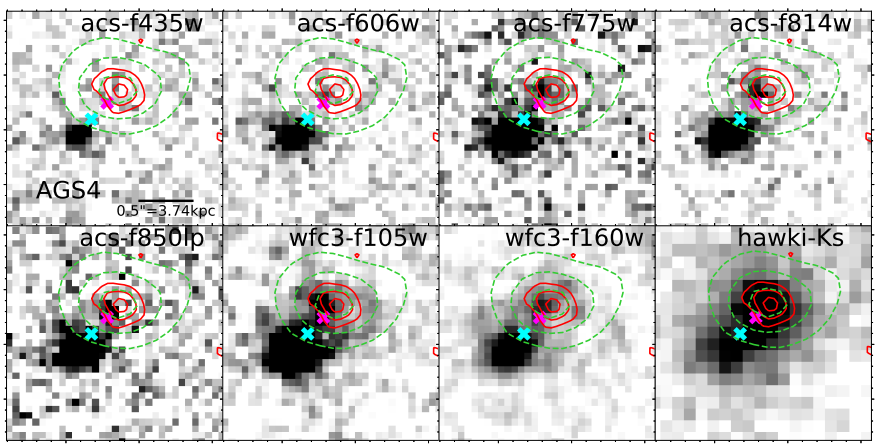}
    \caption{Images of AGS4 from the observed B (435\,nm, rest-frame UV) to near-infrared observed $K_{s}$ (2.2\,$\mu$m, rest-frame B) bands. 
The red contours denote the ALMA detection at 1.1mm at the resolution of 0\farcs29.  
The green dashed contours denote the ALMA detection at 2mm at  the resolution of 0\farcs88.
The astrometric corrections between ALMA and HST images are applied throughout this work.
The cyan "x" denotes the optical bright neighbor, ID$_{\rm CANDELS}$\,=\,8923 ($z_{phot}$\,=\,0.24).
The magenta "x" denotes the  detection by ZFOURGE based on the $K_s$-band image, ID$_{\rm ZFOURGE}$\,=\,12333 ($z_{phot}$\,=\,3.76). If not stated otherwise, the images in this paper are all oriented with the north up and east to the left.
} 
\label{fig:ags4-img}
\end{figure*}

AGS4 is  detected in GOODS-ALMA with a $S/N$\,=\,9.7 at 1.1\,mm (\citetalias{Franco18},  see also red solid contours in Fig.~\ref{fig:ags4-img}).  
Using the ALMA follow-up observations presented in Section~\ref{sec:AGS4line}, we now have a confirmation of the robustness of this source through its detection with ALMA at 2\,mm with an even higher $S/N$\,=\,14.7 (green dashed contours in Fig.~\ref{fig:ags4-img}). 


AGS4 is an illustration of what we may call spectral de-confusion. The two catalogs CANDELS and ZFOURGE have identified a single object associated with this position on the sky.
The CANDELS position is close to the center of the bright optical emitter whereas the ZFOURGE position is close to the bright $K_s$-band near-infrared emitter.
The position of the ALMA detection exhibits a clear offset with respect to the bright optical emitter and falls on the northwestblob that is the brightest in the near-infrared.
As shown in Fig.~\ref{fig:ags4-img}, AGS4 is at 0\farcs5 from a  bright optical emitter. They are considered as two regions of the same galaxy  (ID$_{\rm CANDELS}$\,=\,8923, $z_{phot}$\,=\,0.24,  cyan "x" in Fig.~\ref{fig:ags4-img}) in the CANDELS catalog \citep{Guo13}. The ZFOURGE  position is at 0\farcs16 (ID$_{\rm ZFOURGE}$\,=\,12333)  to AGS4 (magenta "x"  in Fig.~\ref{fig:ags4-img}), and the photometic redshift derived  by the ZFOURGE team is $z^{12333}_{phot}$\,=\,3.76. 

As discussed above, AGS4 exhibits an $H$-band counterpart well above the WFC3 detection limit in GOODS-\textit{South} of $H$\,=\,25.23\,AB that was missed due to blending with a bright neighbor. Therefore, despite being truly optically dark in the optical bands, AGS4 is not strictly speaking an optically dark galaxy since it is detected in the $H$-band. After de-blending\footref{de-blending} both emitters -- none of which corresponds to the centroid of the CANDELS or ZFOURGE positions that are both in between the two emitters --  we can clearly see two different SEDs (Fig.~\ref{fig:sed}-{\it Top-left}). We note that the two ALMA detections at 1.1mm and 2mm are well-centered on the northwestemitter. The photometric redshift associated to this object, once deconvolved from the bright optical emitter, is $z_{phot}^{AGS4}$\,=\,4.02$^{+0.36}_{-0.60}$. 
We note that the optical source identified in the CANDELS catalog (ID$_{\rm CANDELS}$\,=\,8923) is clearly offset from the ALMA contours, leaving little doubt that these are two separate galaxies.
It is only in the low resolution $K_s$-band image that some confusion takes place. Being near-IR based, the ZFOURGE catalog identified the position of the peak $K_s$ emission that matches the ALMA galaxy and attributed to it a photometric redshift of  $z_{phot}^{\rm ZFOURGE}$=3.76 (ID$_{\rm ZFOURGE}$\,=\,12333). 
The SED fittings of the photometric points associated to the CANDELS (gray) and ZFOURGE (green) galaxies are both shown in Fig.~\ref{fig:sed}-{\it Top-left}. The gray SED does not leave any doubt on the fact that this is a very nearby galaxy ($z_{phot}^{\rm CANDELS}$=0.24), with a clear drop of the emission above 0.8\,$\mu$m, whereas the green SED exhibits a break around 1.6\,$\mu$m ($z_{phot}^{\rm ZFOURGE}$=3.76). The very strong difference between the SEDs of the CANDELS and ZFOURGE sources clearly show that they cannot be considered as two sides of the same galaxy, one visible in the optical the other one in the near to far infrared. 
These two redshifts suggest that there are indeed two galaxies in projection here.

Our ALMA spectral-scan follow-up presented in Section~\ref{sec:AGS4line} provides complementary evidence for this spectral de-confusion between two super-imposed galaxies with a clear line detection that matches the CO(6-5) transition at a redshift of $z_{spec}$\,=\,3.566 (see Fig.~\ref{fig:spec}-\textit{Left}), well within the range of possibilities of our photometric redshift estimate. 
Adopting the redshift at  $z_{spec}^{AGS4}$\,=\,3.556, the stellar mass of AGS4 is {\it M}$_\star$\,=\,10$^{11.09\pm0.08}$\,M$_\odot$, which makes  AGS4   one of the most massive galaxies in the early Universe.  We note that  adopting  $z_{spec}^{AGS4}$\,=\,4.326, as for CO(7-6), the stellar mass that we obtain for AGS4 is {\it M}$_\star$\,=\,10$^{11.45\pm0.20}$\,M$_\odot$, which would make AGS4 the most massive galaxies known in the Universe above $z$\,=\,4.

We investigated the possibility that the stellar mass of AGS4 may be contaminated by the presence of the continuum from an AGN, but we did not find any evidence of a power-law continuum emission in the SED presented in the Fig.\,23 of \citetalias{Franco18}  (also in \citetalias{Franco2020b}), nor any radio emission associated to it. There is an X-ray detected AGN that is associated with its neighbor (ID$_{\rm ZFOURGE}$\,=\,12333), which could potentially affect the measurement if attributed to the wrong counterpart, but again we do not see any clear signature of such effect. The {\it q}$_{\rm TIR}$-parameter (see Eq.~\ref{eq:q}) measuring the infrared to radio flux ratio parameter is equal to $q_{\rm TIR}$\,=\,2.49$\pm$0.06, higher than the typical value for star-forming galaxies at $z$\,$\sim$\,3.5 proposed by \citet{Delhaize17},  $q_{\rm TIR}$\,=2.16$\,\pm\,$0.06, indicating no radio excess of this galaxy.

\begin{equation}
\centering
    q_{\rm TIR} = \rm log\left(\frac{{\it L}_{TIR}} {3.75 \times 10^{12} \, Hz }\right)-log\left(\frac{{\it L}_{1.4\,GHz}}{ W \, Hz^{-1}}\right),
	\label{eq:q}
\end{equation}
where {\it L}$_{\rm TIR}$ is total infrared luminosity (8\,--\,1000 $\mu$m)  in unit  W,  and the $L_{ \rm 1.4\,GHz}$\,=\,2.73$\times$10$^{24}$\,W\,Hz$^{-1}$   is derived from the  10.4$\sigma$ detection of AGS4 at 3\,GHz assuming a radio spectral index of $\alpha$\,=\,$\alpha^{\rm 3GHz}_{\rm 1.4GHz}$\,=\,$-0.8$, the typical value for galaxies at $z$\,>\,2 \citep{Delhaize17}, using Eq.~\ref{eq:S1d4GHz}. 

\subsection{AGS25, the most distant optically dark galaxy in GOODS-ALMA} 
\label{sec:AGS25}

\begin{figure}[htbp]
  \centerline{
    \includegraphics[width=\columnwidth]{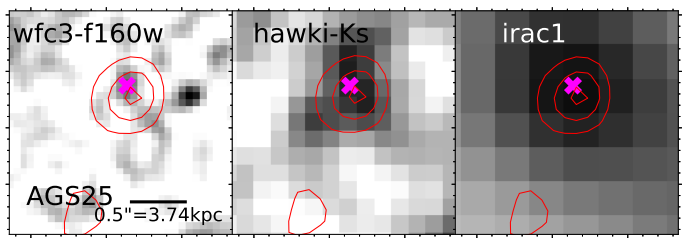}
}
\caption{
Images of AGS25 in the $H$-band (1.6\,$\mu$m), $K_s$-band (2.2\,$\mu$m) and IRAC1-band (3.6\,$\mu$m). 
The contribution of a 3\arcsec distant bright neighbor of AGS25 was subtracted to produce the IRAC image as in \citetalias{Franco2020a}.
The red contours denote the ALMA detection at 1.1mm at the resolution of 0\farcs60.  The magenta "x" is the ZFOURGE counterpart  ID$_{\rm ZFOURGE}$\,=\,11353. 
 }
  \label{fig:AGS25}
  \end{figure}

AGS25 is the only optically dark galaxy in the sample that is identified in the ZFOURGE catalog and is not confused with a neighboring galaxy.
The counterpart of AGS25 in the $K_{\rm s}$-band image is clearly visible in Fig.~\ref{fig:AGS25} (middle panel). It is associated with the ZFOURGE galaxy  ID$_{\rm ZFOURGE}$\,=\,11353 at $z^{\rm AGS25}_{phot}$=4.64. We fitted its photometric measurements as listed in the ZFOURGE catalog using the code \texttt{EAzY} for consistency with the other galaxies, and found a PDF of the photometric redshift well peaked at the redshift given in the ZFOURGE catalog (Fig.~\ref{fig:sed}-\textit{Bottom-right}). We find a peak redshift of $z^{\rm AGS25}_{phot}$\,=\,4.71$^{+0.24}_{-0.24}$ that encompasses the value from ZFOURGE that we keep for consistency with \citetalias{Franco2020a}.
This makes AGS25 the most distant optically dark galaxy in GOODS-ALMA.
The IRAC emission of the galaxy is polluted by the contribution of a bright neighbor at a distance of 3 arcsec (outside of the postage stamps shown in Fig.~\ref{fig:AGS25}). This explains why the galaxy is not listed in the IRAC S-CANDELS catalog \citep{Ashby15}.
After de-convolution of the bright neighbor, a clear IRAC detection is obtained (see Fig.~\ref{fig:AGS25} right panel) that was used to determine its flux density in \citetalias{Franco2020a}. The stellar mass of AGS25 at this redshift is {\it M}$_\star$\,=\,10$^{10.39^{+0.12}_{-0.29}}$\,M$_\odot$ \citepalias{Franco2020a}.
AGS25 is not detected by Herschel, hence its SFR\,=\,839\,$\pm$131\,M$_{\odot}$\,yr$^{-1}$ was derived in \citetalias{Franco2020a} by adjusting the  SED of \cite{Schreiber18} to the 1.1mm ALMA flux density corresponding to 195\,$\mu$m in the rest-frame, which is not far from the peak of the far-infrared SED.

%
%
\section{An overdensity at $z$\,$\sim$\,3.5 in GOODS-ALMA}
\label{sec:overdensity}

\subsection{Clustering properties of optically dark galaxies}
\label{sec:clusterGals}
A large statistical sample of IRAC detected optically dark galaxies in the CANDELS fields  studied by \citet{Wang19} shows that  the optically dark galaxies are the most massive  galaxies in the early Universe and are probably the progenitors of the most massive galaxies we see in the  groups or clusters of galaxy in the local Universe. 

Here in the  69 arcmin$^2$  GOODS-ALMA field, four of the six optically dark galaxies reside in a small region of 5 arcmin$^2$ only. We calculate that the probability to have four out six galaxies randomly falling in such a small area  is only  0.4\%  from four million Monte Carlo mock realizations. In addition, all the four optically dark galaxies present a redshift consistent with $z$\,=\,3.5 where field galaxies demonstrate a clear peak in their redshift distribution.

\subsection{A clear peak at $z$\,$\sim$\,3.5 in the redshift distribution}
\label{sec:zdistribution}
\begin{figure}[htbp]
	\includegraphics[width=\columnwidth]{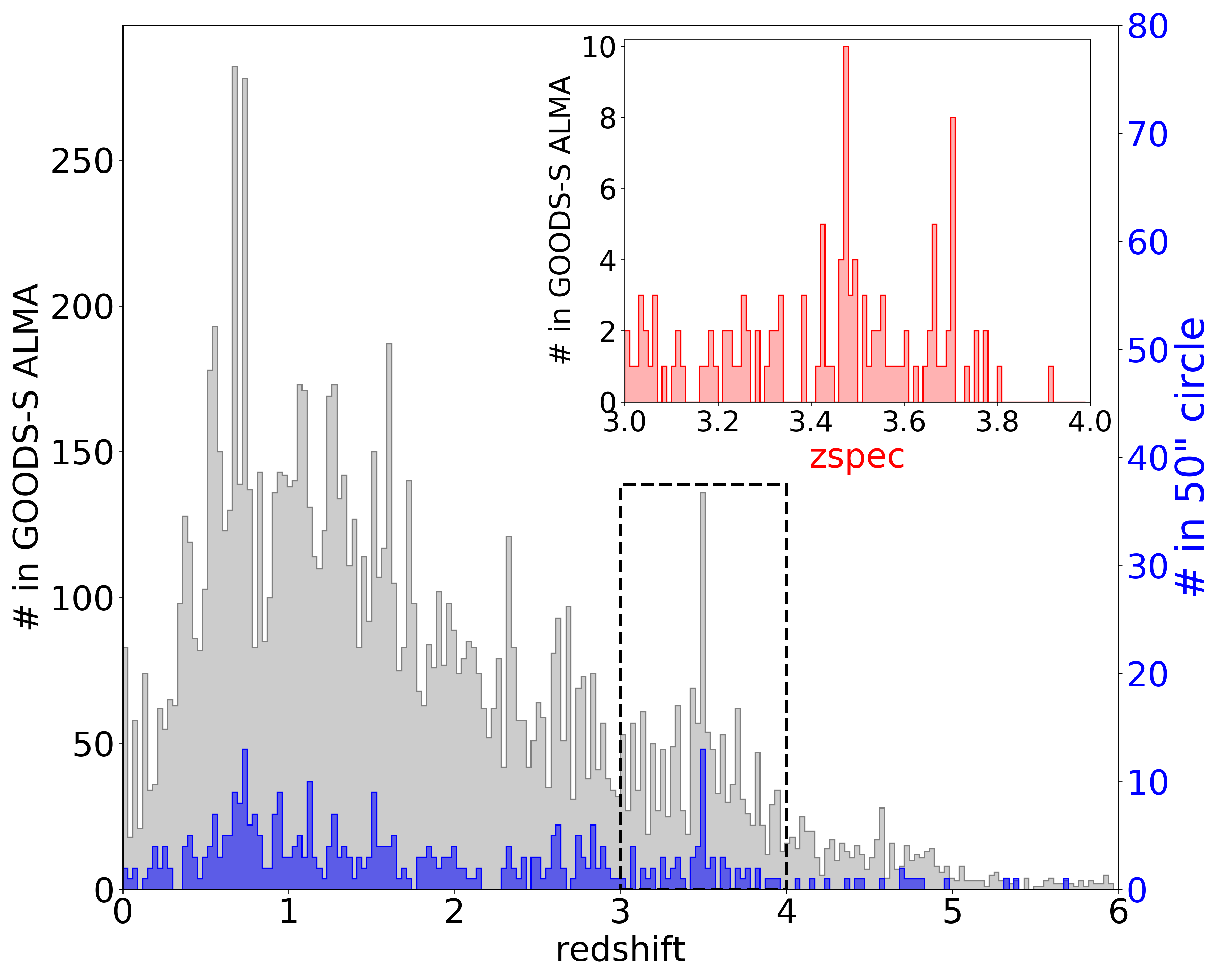}
    \caption{Redshift distribution of the 11674 galaxies with a redshift (either photometric or spectroscopic) located within the 6.9$\arcmin$$\times$10$\arcmin$  GOODS-ALMA field (gray). The subsample of 395 galaxies located within 50$\arcsec$ of the most massive galaxy AGS24, are shown in blue. The insert shows the distribution of the spectroscopic redshifts of the  121 galaxies  at 3\,$\leq$\,$z_{spec}$\,$\leq$\,4 over the GOODS-ALMA field. Ten galaxies fall in the redshift bin 3.470\,$\leq$\,$z_{spec}$\,$\leq$\,3.480. The mean redshift of the ten galaxies is $z_{spec}$\,=\,3.472. The redshifts are from the ZFOURGE \citep{Straatman16} and VANDELS DR3 \citep{McLure18, Pentericci18} catalogs.}
    \label{fig:histogram}
\end{figure}

The redshift distribution of the galaxies with either a photometric or a spectroscopic redshift in the 6.9$\arcmin$$\times$10$\arcmin$ GOODS-ALMA field presents a peak at $z$\,$\sim$\,3.5 (Fig.~\ref{fig:histogram}). This peak is at a 3.5$\sigma$ significance among the 1373 galaxies at 3\,$\leq$\,$z$\,$\leq$\,4 ( Fig.~\ref{fig:histogram}). The redshift bin size is taken to be 0.03 to optimize the peak in the redshift distribution that was previously identified in \citet[Fig. 23]{Straatman16} using the ZFOURGE catalog. 

Here we increase by a factor of three the total number of galaxies with a spectroscopic redshift that fall within the redshift range 3\,$\leq$\,$z$\,$\leq$\,4 in GOODS-ALMA. As many as 83 new spectroscopic redshifts have recently been measured and released in the data release 3 (DR3) of the VANDELS survey \citep{McLure18, Pentericci18} that fall within 3\,$\leq$\,$z$\,$\leq$\,4 in GOODS-ALMA. The VANDELS redshifts supplement the already existing 38 spectroscopic redshifts originally listed in the ZFOURGE catalog and used by \cite{Straatman16}. 


The 121 galaxies with spectroscopic redshifts  at 3\,$\leq$\,$z_{spec}$\,$\leq$\,4 show a clear peak at  3.470\,$\leq$\,$z_{spec}$\,$\leq$\,3.480 containing ten galaxies  (Fig.~\ref{fig:histogram}). These ten galaxies have a mean redshift at  $z_{spec}$\,=\,3.472. This is consistent with the peak at $z$\,$\sim$\,3.5 shown by the 1373 galaxies with either spectroscopic or photometric redshifts.

\subsection{optically dark galaxies at $z$\,$\sim$\,3.5}
\label{sec:physics}

\begin{sidewaystable*}[htbp]
	\centering
	\scriptsize
	\caption{Properties of the optically dark galaxies.}
	\label{tab:hstdark}
	\begin{threeparttable}
	\begin{tabular}{lllllllllllllllllllllllllllll} 
\hline\hline
ID     & RA    &Dec  &$z_{phot}$ &$z_{spec}$   & assumed $z$ &log\,{\it M}$_{\star}$  &log\,{\it L$_{\rm IR}$ } &SFR                                &$R_{\rm SB}$ &{\it F$_{\rm 1.1mm}$} &{\it F$_{\rm 2mm}$} &log\,{\it L$_{\rm CO}$ } &$H$   &{\it F$_{\rm 3GHz}$} &{\it F$_{\rm 6GHz}$}  &{\it L$_{\rm 1.4GHz}$}  \\%
        &[deg]  &[deg] &                  &                     &                       &log\,[M$_{\odot}$]     &log\,[L$_{\odot}$]        &[M$_{\odot}$\,yr$^{-1}$]  &                       &[mJy]                          &[mJy]                        &log\,[L$_{\odot}$]         &[mag] &[$\mu$Jy]                  &[$\mu$Jy]                  &[W\,Hz$^{-1}$]                  \\
\noalign{\vskip 4pt}
(1)&(2)&(3)&(4)&(5)&(6)&(7)&(8)&(9)&(10)&(11)&(12)&(13)&(14)&(15)&(16)&(17)\\
\noalign{\vskip 2pt}
\hline
\noalign{\vskip 3pt}
AGS4      &53.148839 &-27.821192 &-- &3.556 &3.556                            &11.09$^{+0.06}_{-0.18}$       &12.93$\pm$0.02     &1435$^{+76}_{-83}$        &3.8$^{+1.9}_{-0.5}$         &1.72$\pm$0.20                  &0.29$\pm$0.02              &4.95$\pm$0.93      &25.23 &17.28$\pm$1.66  &8.64$\pm$0.77 &2.73$\times$10$^{24}$            \\ 
\noalign{\vskip 2pt}
AGS11    &53.108818 &-27.869055 &--&--  &3.472                                   &10.24$^{+0.75}_{-0.00}$      &12.94$\pm$0.07     &1492$^{+252}_{-253}$    &28.7$^{ +0.1}_{-23.6}$     &1.34$\pm$0.25                  &0.30$\pm$0.02              &--                            &--       &4.48$\pm$1.06    & --                       &6.72$\times$10$^{23}$        \\
\noalign{\vskip 2pt} 
AGS15    &53.074847 &-27.875880 &3.96$^{+0.80}_{-0.78}$&--&3.472  &10.56$^{+0.01}_{-0.41}$      &12.78$\pm$0.08     & 1034$^{+180}_{-179}$   &9.5$^{+11.2}_{ -1.7}$       &1.21$\pm$0.11                  &0.32$\pm$0.02              &--                            &27.11 &5.38$\pm$1.06    & --                       &8.08$\times$10$^{23}$       \\
\noalign{\vskip 2pt}
AGS17    &53.079374 &-27.870770  &--&3.467&3.467                             &10.52$^{+0.40}_{-0.06}$      &13.08$\pm$0.02    & 2070$^{+112}_{-117}$   & 20.9$^{ +3.1}_{-12.6}$    &2.30$\pm$0.44                  &0.35$\pm$0.02              &5.22$\pm$1.01      & --      &39.00$\pm$3.94  &1.16$\pm$0.32  &5.86$\times$10$^{24}$          \\
\noalign{\vskip 2pt} 
AGS24  &53.087178 &-27.840217  &3.58$^{+0.49}_{-0.38}$&--& 3.472  &11.32$^{+0.02}_{-0.19}$     &12.31$\pm$0.11     &353$^{+80}_{-82}$         &0.6$^{+0.5}_{-0.1}$          &0.88$\pm$0.22                  &0.19$\pm$0.02              &--                            &--       &12.43$\pm$2.19  &1.41$\pm$0.32  &1.87$\times$10$^{24}$              \\ 
\noalign{\vskip 2pt}
AGS25  &53.183710 &-27.836515   &4.64 &-- & 4.64                               &10.39                                    &12.68$\pm$0.07     &832$^{+126}_{-135}$     &8.0$^{+1.3}_{-1.3}$          &0.82$\pm$0.19                  &--                                    &--                            & --      &3.92$\pm$1.06    &1.07$\pm$0.32  &5.90$\times$10$^{23}$                 \\
\noalign{\vskip 2pt}
\hline
	\end{tabular}
        \begin{tablenotes}  
        \item Columns: \textbf{(1)} Source names;  \textbf{(2)}  \textbf{(3)} Coordinates in the ALMA image (J2000): see \citetalias{Franco18} and  \citetalias{Franco2020a}; \textbf{(4)} Photometric redshifts: see Section~\ref{sec:z};  \textbf{(5)} Spectroscopic redshifts: The spectroscopic redshifts of AGS4 and AGS17 are  derived from the CO(6-5) line measured in our ALMA spectroscopic scan follow-up, see Section \ref{sec:Lines}; \textbf{(6)} Assumed redshifts: We assign the redshift of AGS11, AGS15 and AGS24 to be the same as the central position of the redshift peak that presents an overdensity centered on AGS24, $z_{peak}$\,=\,3.472, as discussed in section~\ref{sec:zdistribution} and \ref{sec:physics}, which is consistent with their photometric redshifts. For each galaxy, all the properties are measured assuming the redshift listed here; \textbf{(7)} Stellar masses: see Section~\ref{sec:mstar}; \textbf{(8)} Infrared luminosities: see Section~\ref{sec:lir}; \textbf{(9)} SFRs: see Section~\ref{sec:sfr}; \textbf{(10)} Starburstness: As in \citetalias{Franco2020a}, $R_{\rm SB}$\,=\,SFR/SFR$_{MS}$, where  SFR$_{MS}$ is the average SFR of MS galaxies as defined in \cite{Schreiber18}; \textbf{(11)}  Flux densities at 1.1\,mm: Peak fluxes measured using \texttt{Blobcat} as in \citetalias{Franco18} and the fluxes of AGS15 and AGS17 are updated in \citetalias{Franco2020a}; \textbf{(12)} Peak fluxes at 2\,mm from the ALMA band4 follow-up; \textbf{(13)} CO luminosities: the CO(6-5) line detected in our ALMA spectroscopic scan follow-up; \textbf{(14)} $H$-band AB magnitudes:  After de-blending, AGS4 and AGS15 turn out to have $H$-band magnitudes higher than the detection limit; \textbf{(15)} Flux densities at 3GHz: the flux densities of AGS4, 17 and 24 are the peak flux  derived using PyBDSF\footref{pybdsf}(Rujopakarn et al. in prep.), flux densities of AGS11, 15 and 25 are the peak flux densities in the 3GHz image; \textbf{(15)} Flux densities at 6\,GHz: the flux density of AGS4 is the peak flux derived using PyBDSF\footref{pybdsf},  flux densities of AGS17, 24 and 25 are the peak flux densities in the 6GHz image, AGS11 and AGS15 do not have peak flux density higher than 3$\sigma$; \textbf{(17)} Luminosity at 1.4GHz: derived from the flux density at 3GHz using Eq.~\ref{eq:S1d4GHz}.
       \end{tablenotes}
    \end{threeparttable}
\end{sidewaystable*}

We find that four out of the six optically dark galaxies exhibit a redshift that is consistent with being located in the same redshift peak at $z$$\sim$3.5, which corresponds to an overdensity centered on the most massive of them, AGS24, as to be discussed below.  
We may even count five out of the six if we include AGS4, which is at $z_{spec}$\,=\,3.556 and at a distance of 3$\arcmin$  from AGS24. Accounting for the redshifts of the sources, this corresponds to 70.8 cMpc (comoving) and 15.5 pMpc (proper). The comoving distance is typical of those found in distant super-clusters while the proper distance is consistent with those of proto-clusters, which suggests that AGS4 could potentially be a member of the proto-cluster as well.
We start by describing in detail the properties of AGS24, because it is the most massive of all the optically dark galaxies in GOODS-ALMA and even among all GOODS-\textit{South} galaxies at $z$\,$>$\,3 and constitutes an excellent candidate for a  future BCG galaxy. The remaining three optically dark galaxies at $z$$\sim$3.5 are discussed in the next section.

\subsubsection{AGS24,  the most massive galaxy at $z$\,$>$\,3 in GOODS-{\it South}  } 
\label{sec:massivegal}
  \begin{figure*}[htbp]
  \centerline{
 	\includegraphics[height=2.3in]{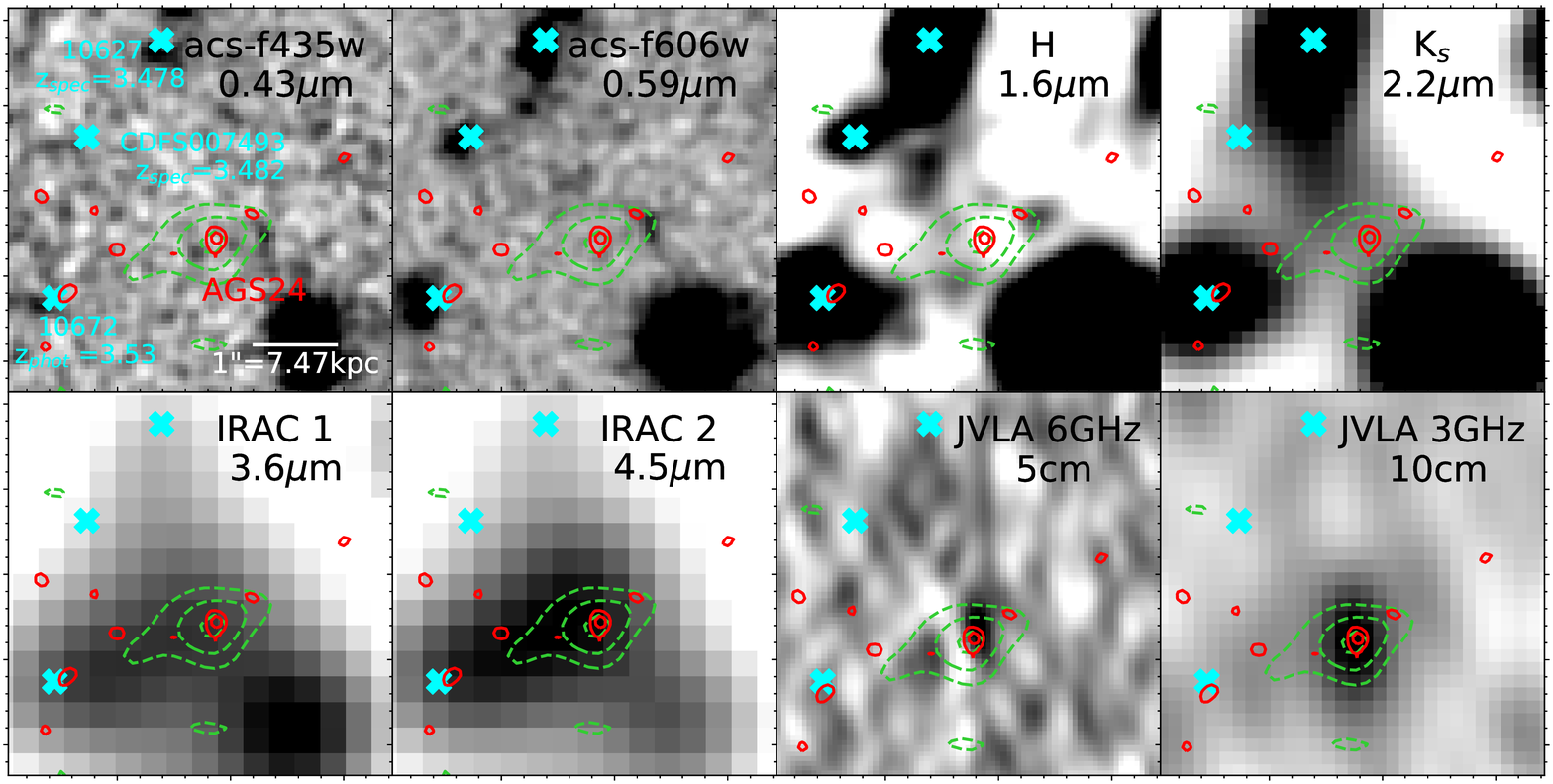}
}
\caption{
Images of AGS24 in the {\it B}, {\it V},  {\it H} and {\it K$_{s}$} band, IRAC 3.6 and 4.5$\mu$m, JVLA 6 GHz (3.7$\sigma$) and 3 GHz (5.7$\sigma$). 
The contours are the same as in Fig.~\ref{fig:ags4-img} but for AGS24.
The three neighbor galaxies are labeled with ID$_{\rm ZFOURGE}$ or ID$_{\rm CDFS}$ and their redshift separately.
 }
\label{fig:img_AGS24}
  \end{figure*}

\begin{table}
\centering
\footnotesize
\caption{Information of the neighbors of AGS24.}
\label{tab:neighbors}
\begin{tabular}{lrrrrrrr} 
\hline\hline
ID  &  \it z$_{spec}$  &\it z$_{phot}$     &   {\it M$_{\star}$}   &SFR   \\
&&&[10$^{10}\,\times$\,M$_{\odot}$]&[M$_{\odot}$\,yr$^{-1}$]\\
\hline
ZFOURGE10672 &    --      &3.53 &5.88 &30.2\\
ZFOURGE10627 & 3.478  &3.48 &2.63  &14.8\\
CDFS007493       & 3.482  & -- &0.41  &--\\
\hline
\end{tabular}
\end{table}

AGS24 has been detected with a S/N of 4.9$\sigma$ and 8.7$\sigma$ in the GOODS-ALMA 1.1mm image and in the 2\,mm continuum image of the ALMA spectroscopic scan follow-up (red and green contours in Fig.~\ref{fig:img_AGS24}).
This galaxy exhibits a lower S/N  in the GOODS-ALMA image tapered at 0\farcs6 (3.9$\sigma$) than in the image at the original resolution of 0\farcs29 (4.9$\sigma$), which suggests that the dust emission is very compact.
This optically dark galaxy exhibits a clear radio counterpart at 6 GHz (3.7$\sigma$) and 3 GHz (5.7$\sigma$).  
Despite being absent from the list of IRAC sources in the field given in the S-CANDELS catalog \citep{Ashby15}, this source does exhibit clear IRAC emission at 3.6 and 4.5\,$\mu$m that is partly blended with the three surrounding very nearby galaxies that are clearly detected in the {\it H} and {\it K$_{s}$} bands. The emission of AGS24 in the {\it H} and {\it K$_{s}$} bands is marginal but was estimated after de-blending the contribution from the three neighboring sources and used to determine its photometric redshift. 

After de-blending\footref{de-blending}, we derive a photometric redshift of $z_{\rm AGS24}$\,=\,3.58$^{+0.49}_{-0.38}$. Assuming this redshift, the excess emission in the {\it V} band may correspond to the Ly$\alpha$ emission line (Fig.~\ref{fig:sed}-\textit{Bottom-left}).  However, we analyzed the VLT/MUSE data obtained in this region but did not find any evidence for Ly$\alpha$ emission despite the high sensitivity of MUSE \citep[$rms$\,=\,4$\times$10$^{20}$\,erg\,s$^{-1}$\,cm$^{-2}$\,A$^{-1}$][]{Herenz17}.  
We conclude that the excess $V$ band emission remains uncertain and may be due to noise fluctuation.

The three neighboring galaxies are located at similar distances of $\sim$1\farcs5 (equivalent to 11.3 kpc at $z$\,$\sim$\,3.5) from AGS24 (Fig. \ref{fig:img_AGS24} and Table \ref{tab:neighbors}). They all have a redshift of $z$$\sim$3.5, including two spectroscopic redshifts  -- ID$_{\rm ZFOURGE}$\,=\,10627, $z$$_\textit{spec}$\,=\,3.478 and ID$_{\rm CDFS}$\,=\,007493, $z$$_\textit{spec}$\,=\,3.482 (VANDELS DR3) -- and one photometric redshift -- ID$_{\rm ZFOURGE}$\,=\,10672, $z$$_{phot}$\,=\,3.53 \citep{Straatman16}. As a result, the group of four galaxies appears to be physically connected and makes a dense and compact association of galaxies typical of what can be found at the center of distant proto-clusters  \citep{Springel05, Lucia07}.

Altogether, the following reasons suggest that AGS24 is located at the redshift of $z$$\sim$3.472 (Section~\ref{sec:zdistribution}) of the proto-cluster: \textit{(i)} the three neighbors of AGS24 have $z$$\sim$3.472, \textit{(ii)} the photometric redshift of AGS24 is consistent with this redshift, \textit{(iii)} AGS24 belongs to the highly improbable association of optically dark galaxies that fall in the region of the $z$$\sim$3.472 proto-cluster. 

Assuming this redshift, we estimate the total infrared luminosity of AGS24 to be $L_{\rm IR}$\,=\,10$^{12.31\pm0.11}$\,L$_{\odot}$ fitting the rest-frame 195, 245, 490\,$\mu$m emission of the galaxy that corresponds to the observed 870\,$\mu$m, 1.13 and 2 mm using the typical far-infrared SED of a MS galaxy at this redshift \citep{Schreiber15}. We also checked that we obtained a consistent total infrared luminosity after fitting these measurements with a \citet{Draine2007} SED, which gives the same result within 10\%.

Assuming $z$\,=\,3.472, the stellar mass of AGS24 is {\it M$_{\star}$}\,=\,10$^{11.32^{+0.02}_{-0.19}}$\,M$_{\odot}$ (adopting the \citealt{Calzetti00} attenuation law and a delayed exponentially declining SFH). It is $\sim$\,10 times more massive than the galaxy presented in \cite{Ginolfi17}, which used to be the most massive galaxy at $z$\,=\,3$\sim$\,4 in this field. We note that the stellar mass was derived by fitting the near UV to near-infrared photometry, similar to the method we used for our galaxies. In order to test the robustness of the value of the stellar mass of AGS24, we recompute it assuming various star formation histories (SFH)   to see whether this mass could fall down by a large amount. We use the SED fitting code  \texttt{FAST++}\footref{FAST++} to test a delayed exponentially declining and an exponentially declining SFH as well as a truncated SFH, with the possibility to add a second burst. We also test the impact of varying the attenuation law using the code  \texttt{CIGALE} \citep{Boquien19} with the attenuation laws from \citet{Calzetti00}, \citet{Charlot00}, and the modified \citet{Charlot00} law in MAGPHYS \citep{daCunha08}. The results are listed in Table~\ref{tab:SEDmodels}. They show that despite these wide variety of SFH and attenuation laws, the stellar mass of AGS24 remains within 10$^{11.31}$ and 10$^{11.52}$\,M$_{\odot}$.

With a redshift of $z$\,$\sim$\,3.5, AGS24 is the most massive galaxy in the GOODS-ALMA field above $z$\,=\,3. The other  similarly massive candidate galaxies in this redshift range either have a strong AGN potentially contaminating the estimate of the stellar mass or highly unreliable redshifts.
Two galaxies without an AGN at $z$\,>\,3 have similarly large stellar masses in the GOODS-ALMA field (ID$_{\rm ZFOURGE}$\,=\,11505 and 16410, marginally detected in the $K_s$-band with  S/N\,=\,5.0, 5.5, at $z_{phot}$\,=\,5.36, 4.73). 
They both have very high $\chi^2$ values of 126.6 and 43.0 associated with their photometric redshifts. Only 3\% and 19\% of the ZFOURGE galaxies have  photometric redshifts with such high $\chi^2$ values.
Therefore, we consider the redshifts of these two galaxies not reliable.

We inspect the possible contamination of an AGN on the near- and mid-infrared emission, which may result in an overestimation of the stellar mass. We calculate the infrared to radio flux ratio, $q_{\rm TIR}$\,=\,2.05$\pm$0.18 (see Eq.~\ref{eq:q}), where the 1.4GHz luminosity, $L_{ \rm 1.4\,GHz}$\,=\,1.87$\times$10$^{24}$\,W\,Hz$^{-1}$, is converted from the JVLA 3GHz emission (5.7$\sigma$) of AGS24 assuming a radio spectral index of $\alpha$\,=\,$-0.8$, as we did for AGS4. Within the rms, the infrared to radio flux ratio agrees  with the  value  $q_{\rm TIR}$\,= 2.16$\,\pm\,$0.06 for star-forming galaxies at redshift $z$\,$\sim$\,3.5  by \citet{Delhaize17}, indicating no radio excess of this galaxy. We also checked that AGS24 does not have a counterpart in the deepest Chandra 7Ms source catalog \citep{Luo17}. We searched for extended diffuse emission in the Chandra 7Ms image but found no robust evidence for such emission that could be associated with an intra-cluster medium. 
Therefore, we conclude that AGS24 is the most massive galaxy at $z$\,>\,3 without an AGN in the GOODS-ALMA field.
\citet{Forrest2019} recently spectroscopically confirmed ($z_{spec}$\,=\,3.493)  a similarly extremely massive quiescent galaxy in the VIDEO XMM-Newton field (Annunziatella et al., in prep.) that is spectroscopically confirmed ($z_{spec}$\,=\,3.493) and interpreted as experiencing a rapid downfall of star formation. 


The predicted number of galaxies that are as massive or more massive than AGS24 in the GOODS-ALMA field at 3\,<\,$z$\,<\,4.5 is 1.15,  based on the galaxy stellar mass function (SMF) derived from the observations in the COSMOS field \citep{Davidzon17}.  \citet{Pillepich2018} compared the SMF of \citet{Davidzon17} to the one  from the IllustrisTNG simulation and  found them consistent. Thus the existence of one massive galaxy, AGS24, is consistent with the existing observation and simulation.
\begin{table}
\centering
\footnotesize
\caption{Stellar masses of AGS24 assuming different SFHs and attenuation laws. }
\label{tab:SEDmodels}
\begin{threeparttable}
\begin{tabular}{lrrrrrrr} 
\hline\hline
\noalign{\vskip 3pt}
\multicolumn{2}{c}{ \texttt{FAST++}\footref{FAST++}:}                                 \\   
\multicolumn{2}{c}{ \citet{Calzetti00} attenuation law + vary SFHs }                                 \\   
\hline
\noalign{\vskip 3pt}
Models            &   log\,{\it M}$_{\star}$   [log\,M$_{\odot}$]           \\
\noalign{\vskip 3pt}
delayed exponentially declining&  11.37$^{+0.01}_{-0.23}$ \\
\noalign{\vskip 2pt}
exponentialy declining &  11.31$^{+0.11}_{-0.19}$   \\
\noalign{\vskip 2pt}
truncated &  11.40$^{+0.00}_{-0.28}$     \\ 
\noalign{\vskip 2pt}
a second burst &  11.33$^{+0.04}_{-0.18}$    \\ 
\noalign{\vskip 2pt}
\hline
\noalign{\vskip 3pt}
\multicolumn{2}{c}{ \texttt{CIGALE}:}                                 \\   
\multicolumn{2}{c}{ delayed exponentially declining SFH + vary attenuation laws}                                \\   
\hline
\noalign{\vskip 3pt}
Models            &   log\,{\it M}$_{\star}$    [log\,M$_{\odot}$]            \\
\noalign{\vskip 3pt}
\citet{Calzetti00} &  11.52$\pm$0.57    \\
\noalign{\vskip 2pt}
\citet{Charlot00}&  11.41$\pm$0.21   \\
\noalign{\vskip 2pt}
MAGPHYS\tnote{a} &  11.44$\pm$0.22 \\
\noalign{\vskip 2pt}
\hline
\end{tabular}
	\scriptsize
    \begin{tablenotes}
        \item[a] A modified \citet{Charlot00} law that is used in MAGPHYS  \citep{daCunha08}
    \end{tablenotes}
\end{threeparttable}
\end{table}

  \begin{figure*}[htbp]
  \centering
	\includegraphics[height=2.3in]{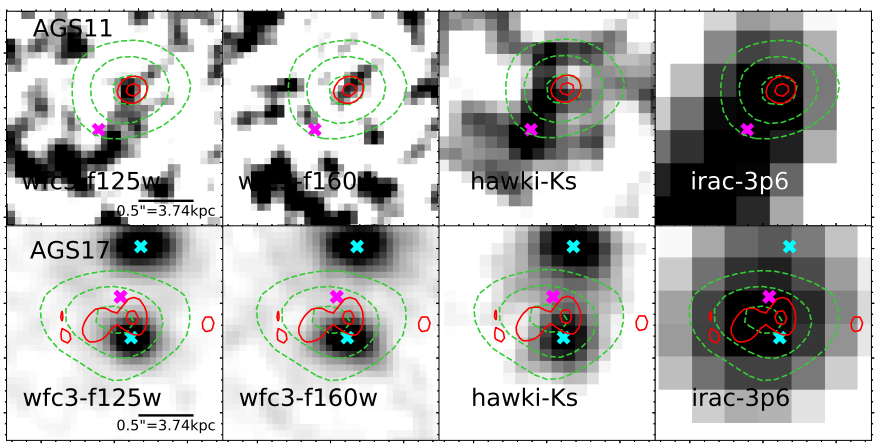}
\caption{
\textbf{\textit{Top:}} Images of AGS11 in the {\it J} band, {\it H} band, {\it K$_{s}$} band and IRAC 3.6\,$\mu$m .
The magenta "x" denotes  the detection in $K_s$-band by ZFOURGE,  ID$_{\rm ZFOURGE}$\,=\,7589 ($z_{phot}$\,=\,4.82).
\textbf{\textit{Bottom:}} Same as above, but for AGS17. 
The cyan "x"s denote the optical bright neighbors, ID$_{\rm CANDELS}$\,=\,4414 ($z_{phot}$\,=\,0.03, below) and ID$_{\rm CANDELS}$\,=\,4436 ($z_{phot}$\,=\,0.95, above).
The magenta "x" denotes the detection in $K_s$-band by ZFOURGE,  ID$_{\rm ZFOURGE}$\,=\,6964 ($z_{phot}$\,=\,1.85).
 }
  \label{fig:AGS1117}
  \end{figure*}

  \begin{figure*}[htbp]
  \centerline{
   	\includegraphics[height=2.3in]{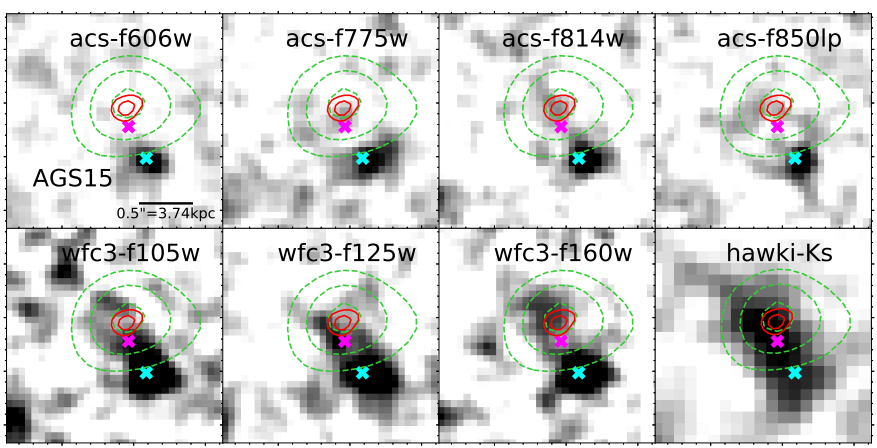}
}
\caption{
Same as in Fig. \ref{fig:ags4-img}, but for AGS15.
The cyan "x" denotes the optical bright neighbor, ID$_{\rm CANDELS}$\,=\,3818 ($z_{phot}$\,=\,4.27).
The magenta "x" denotes the detection in $K_s$-band by ZFOURGE,  ID$_{\rm ZFOURGE}$\,=\,6755 ($z_{phot}$\,=\,3.47).
 }
  \label{fig:AGS15}
  \end{figure*}
%
\subsubsection{Physical properties of the optically dark galaxies AGS11, AGS15 and AGS17 at $z$$\sim$3.5} 
\label{sec:opticaldark}

\paragraph{AGS11:} AGS11 is detected at 1.1mm with a $S/N$\,=\,5.71 in \citetalias{Franco18} (also see red solid contours in Fig.~\ref{fig:AGS1117}-{\it Top}) and 2\,mm with a $S/N$\,=\,16.8 (green dashed contours in Fig.~\ref{fig:AGS1117}-{\it Top}). 
A neighboring galaxy is identified in ZFOURGE that matches the $K_s$ and IRAC emission (magenta "x" in Fig.~\ref{fig:AGS1117}-{\it Top}). This object is located  at $0\farcs$45 from the ALMA detection. It has been attributed with a photometric redshift of $z^{\rm 7589}_{phot}$\,=\,4.82 in the ZFOURGE catalog. However, this redshift is highly uncertain with flag \texttt{use\,=\,0},  hence highly unreliable. Data at hand do not allow us to determine whether this object is the counterpart of the ALMA detection, AGS11, or like several other optically dark sources studied here, an independent galaxy seen in projection. The optical emission at the ALMA detection position is within the noise fluctuation. 
Therefore, we do not  attributed the ZFOURGE ID and the associated redshift to AGS11. Instead, we consider the association of the optically dark galaxies with the overdensity at $z$\,$\sim$\,3.5 as potential evidence that AGS11 has a redshift matching this overdensity.  In Table~\ref{tab:hstdark} we decided to attribute AGS11 with the $z_{peak}$\,=\,3.472 of the overdensity. 
Assuming $z_{\rm AGS11}$\,=\,3.472, the infrared luminosity we derived for AGS11 is  {\it L}$_{\rm IR}$\,=\,10$^{12.94\pm0.07}$\,L$_\odot$. 

\paragraph{AGS15:} AGS15 is located at 0\farcs59  to a optically bright galaxy (ID$_{\rm CANDELS}$\,=\,3818, cyan "x" in Fig.~\ref{fig:AGS15}). It is detected at 5.22$\sigma$ in the 1\,mm image  (\citetalias{Franco18}, see also red contours in Fig.~\ref{fig:AGS15}).  This detection is further confirmed in the 2\,mm continuum   at  15.3$\sigma$  in the follow up ALMA spectroscopic scan (green dashed contours in Fig.~\ref{fig:AGS15}). In the $K_s$-band image, the infrared emission from AGS15 and from the optically bright galaxy is blended, and the two galaxies are confused as one   (ID$_{\rm ZFOURGE}$\,=\,6755, magenta "x" in Fig.~\ref{fig:AGS15}) in the ZFOURGE catalog. The  redshift derived by the ZFOURGE team is $z^{\rm 6755}_{phot}$\,=\,3.46 and the stellar mass is {\it M}$_\star$\,=\,10$^{9.86}$\,M$_\odot$. We notice that  there is a high uncertainty in the SED fitting by the ZFOURGE team,  with  $\chi$$^2$\,=\,45.4, which is higher than 84\% of all the  galaxies in the ZFOURGE catalog. If it is the counterpart of AGS15, then the low stellar mass will make AGS15 an extreme case, that is, ten times less massive compared to the other ALMA detections. 
After de-blending AGS15 and the optically bright galaxy,  as shown in Fig.~\ref{fig:AGS15} and Fig.~\ref{fig:sed}-\textit{Top-right}, we find that in fact AGS15 has an $H$ band magnitude higher than the detection limit, $H$\,=\,27.11\,AB.
We fit the SED with \texttt{EAzY} \citep{Brammer08}  and  \texttt{FAST++}\footref{FAST++} to derive the photometric redshift and  the stellar mass.  The $z_{\rm peak}$ derived for AGS15 is at $z^{\rm AGS15}_{phot}$\,=\,3.96$^{+0.80}_{-0.78}$. Hence the galaxy has, within the uncertainties on the photometry, a photometric redshift that encompasses the redshift peak at $z_{peak}$\,=\,3.472.
Assuming this redshift,  we obtain the infrared luminosity of AGS15 to be {\it L}$_{\rm IR}$\,=\,10$^{12.78\pm0.08}$\,L$_\odot$, the SFR to be SFR\,=\,1042$^{+181}_{-183}$\,M$_{\odot}$\,yr$^{-1}$ and the stellar mass to be {\it M}$_\star$\,=\,10$^{10.56^{+0.01}_{-0.41}}$\,M$_\odot$.  

\paragraph{AGS17:}  AGS17 is detected at 1.1\,mm with a $S/N$\,=\,5.01  (\citetalias{Franco18}, also see red solid contours in Fig.~\ref{fig:AGS1117}-{\it Bottom}) and with a $S/N$\,=\,15.9 in the 2\,mm continuum (green dashed contours in Fig.~\ref{fig:AGS1117}-{\it Bottom}).  
It is located between two bright optical emitters, ID$_{\rm CANDELS}$\,=\,4414, \citep[$z^{4414}_{phot}$\,=\,0.03,][]{Pannella15} at 0\farcs27, and ID$_{\rm CANDELS}$\,=\,4436, ($z^{4436}_{phot}$\,=\,0.95, fitted with \texttt{EAzY}) at 0\farcs57 (cyan "x"s in Fig.~\ref{fig:AGS1117}-{\it Bottom}). These two emitters are also clearly seen in the $K_s$-band image, however,  they are not resolved in the ZFOURGE catalog (magenta "x"  in Fig.~\ref{fig:AGS1117}-{\it Bottom}, ID$_{\rm ZFOURGE}$\,=\,6964). The redshift derived from this blended emission  is $z^{6964}_{phot}$\,=\,1.85, which is not consistent with either the redshifts of the optical emitters. 
 
Assuming that the emission line detected in our ALMA spectroscopic follow-up is the CO(6-5) transition as discussed in Section~\ref{sec:AG17line}, AGS17 has a spectroscopic redshift $z^{\rm AGS17}_{spec}$\,=\,3.467, which puts it at a distance of 4.30\,cMpc from AGS24, assuming $z^{\rm AGS24}$\,=\,$z_{peak}$\,=\,3.472. Hence it falls inside the proto-cluster as the distance is shorter than the typical size of a proto-cluster ($\sim$14.5\,cMpc) at $z$\,$\sim$\,3.5  in simulations \citep{Muldrew2015}.  The low probability to have a galaxy fall in projection with the structure that is located at the  redshift corresponding to the CO(6-5) transition makes us favor this transition.
Then the stellar mass is  {\it M}$_{\star}$\,=\,10$^{10.59^{+0.29}_{-0.19}}$\,M$_\odot$ based on the IRAC band emission and the infrared luminosity derived from the SED fitting using Herschel and ALMA data  is  {\it L}$_{\rm IR}$\,=\,10$^{13.08\pm0.02}$\,L$_\odot$.

\subsection{Spatial distribution of galaxies at $z$\,$\sim$\,3.5 in the GOODS-ALMA field}
\label{sec:clustering}

\subsubsection{Local overdensity ($\delta$) of individual galaxies}
We calculate the overdensity ($\delta$\,=\,$\rho$\,/\,$\Bar{\rho}$) of the 364 galaxies at 3.42\,$\leq$\,$z$\,$\leq$\,3.57 in GOODS-ALMA.  This redshift range covers a distance of 125\,cMpc, which is consistent with the size of a large scale structure,  a (proto-)supercluster of $\sim$\,60\,$\times$\,60\,$\times$\,150\,cMpc  at $z$\,$\sim$\,2.45, identified by \citet{Cucciati2018}. We derive a dispersion of  $\sigma(\delta)$\,=\,0.20 among the 364  galaxies. 
We do that by comparing the  projected local galaxy density within 2\,cMpc ($\rho$) of a galaxy, to the average number density over the GOODS-ALMA field ($\Bar{\rho}$\,=\,1.32\,cMpc$^{-2}$ or 5.13\,arcmin$^{-2}$).


Seven of the detections in the GOODS-ALMA field reside in  this narrow redshift range, AGS1,  \underline{4}, 5, \underline{11},  \underline{15},  \underline{17},  and \underline{24}, five (underlined) out of the seven are  optically dark. The $z^{\rm AGS1}_{spec}$\,=\,3.442 of AGS1 comes from VANDELS DR3,  zflag\,=\,1,  meaning 50\% probability to be correct, while we note that it is recorded as $z^{\rm AGS1}_{spec}$\,=\,2.309 in \citetalias{Franco18}. AGS5 has a photometric redshift of $z^{\rm AGS5}_{phot}$\,=\,3.46 \citep{Straatman16}.  As mentioned in Section~\ref{sec:AGS4}, AGS4 has a redshift of $z_{spec}^{AGS4}$\,=\,3.556. 


At the same time, eight massive galaxies fall in this redshift range in the ZFOURGE catalog. They have a stellar mass, log\,[M$_{\star}$/M$_{\odot}$]\,>\,10.5, which is equivalently massive as  the ALMA detections \citepalias{Franco18}. They have a median stellar mass of {\it M}$_\star$\,=\,10$^{10.82}$\,M$_\odot$.
These eight massive galaxies, as well as the seven ALMA detections  all show a local overdensity ($\delta$\,>\,1). Out of the 15 massive galaxies, AGS24 shows the most prominent clustering, $\delta_{\rm AGS24}$\,=\,2.06, hence a  >\,5$\sigma$ excess with respect to the average galaxy surface density.  Hence AGS24 is both the most massvie galaxy at $z$\,>\,3 in the GOODS-ALMA field with no AGN and it is located in the densest surface density peak at $z$\,$\sim$\,3.5. This makes it an excellent candidate for a future BCG as we discuss in Section \ref{sec:gal}. The degree of overdensity of the four other HST-dark galaxies are $\delta_{\rm AGS11}$\,=\,1.46 (>\,2$\sigma $ excess), $\delta_{\rm AGS15}$\,=\,1.64 (>\,3$\sigma $ excess), $\delta_{\rm AGS17}$\,=\,1.76 ($\sim$\,4$\sigma $ excess), $\delta_{\rm AGS4}$\,=\,1.39 ($\sim$\,2$\sigma $ excess),  and for the other two ALMA detection with optical counterparts, it is  $\delta_{\rm AGS1}$\,=\,1.21, $\delta_{\rm AGS5}$\,=\,1.03.

Regardless of mass,  when focusing on the galaxies in the densest regions, that is to say, $\delta$\,>\,5$\sigma$($\delta$), we found in total 12 galaxies in the ZFOURGE catalog. It turns out that they all lie within 50\arcsec of AGS24 (red circle in Fig. \ref{fig:density_map}) and the barycenter of these  galaxies falls exactly at AGS24. The most massive  of the 12 galaxies is ID$_{\rm ZFOURGE}$\,=\,10672, {\it M}$_\star$\,=\,10$^{10.76}$\,M$_\odot$. It is at 1\farcs5 from AGS24. 
\begin{figure}[htbp]
\centering
	\includegraphics[height=2.6in]{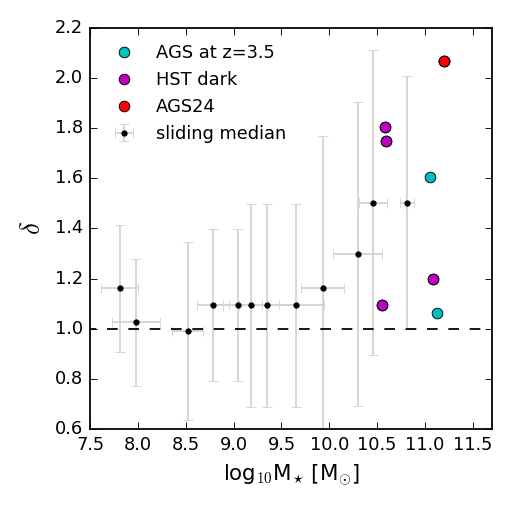}
    \caption{Local overdensity ($\delta$) of individual galaxies in the GOODS-ALMA field at 3.42\,<\,$z$\,<\,3.57. The black dots show the sliding medians of $\delta$ at different stellar masses. The filled circles indicate all the ALMA detections. The purple circles are the HST dark galaxies and the red circle indicates AGS24.}
    \label{fig:deltarho}
\end{figure}

In Fig. \ref{fig:deltarho}, we show the overdensity as a function of stellar mass of the 364 galaxies at 3.42\,<\,$z$\,<\,3.57 in the GOODS-ALMA field and the ALMA detected galaxies. We can see that as the stellar mass increases, the galaxies tend to fall in denser regions, which is in line with what  \citet{Muldrew2015} found in Millennium Simulation \citep{Springel05} on galaxies at $z$\,=\,2. They conclude that the most massive galaxies at high redshift reside in proto-clusters given  the little difference in the environments of only proto-cluster galaxies and those of all the massive galaxies.  Besides, the ALMA detections are the most massive ones in the field and reside in the densest regions. AGS24, as the most massive one in the field, resides in the densest region in the field. Therefore, these optically dark galaxies revealed by ALMA make up for what has been missed from optical observations. The relation between their dusty and massive nature and the dense environment they reside in provides evidence of environmental effects on galaxy evolution.

\begin{figure*}[htbp]
	\includegraphics[width=\hsize]{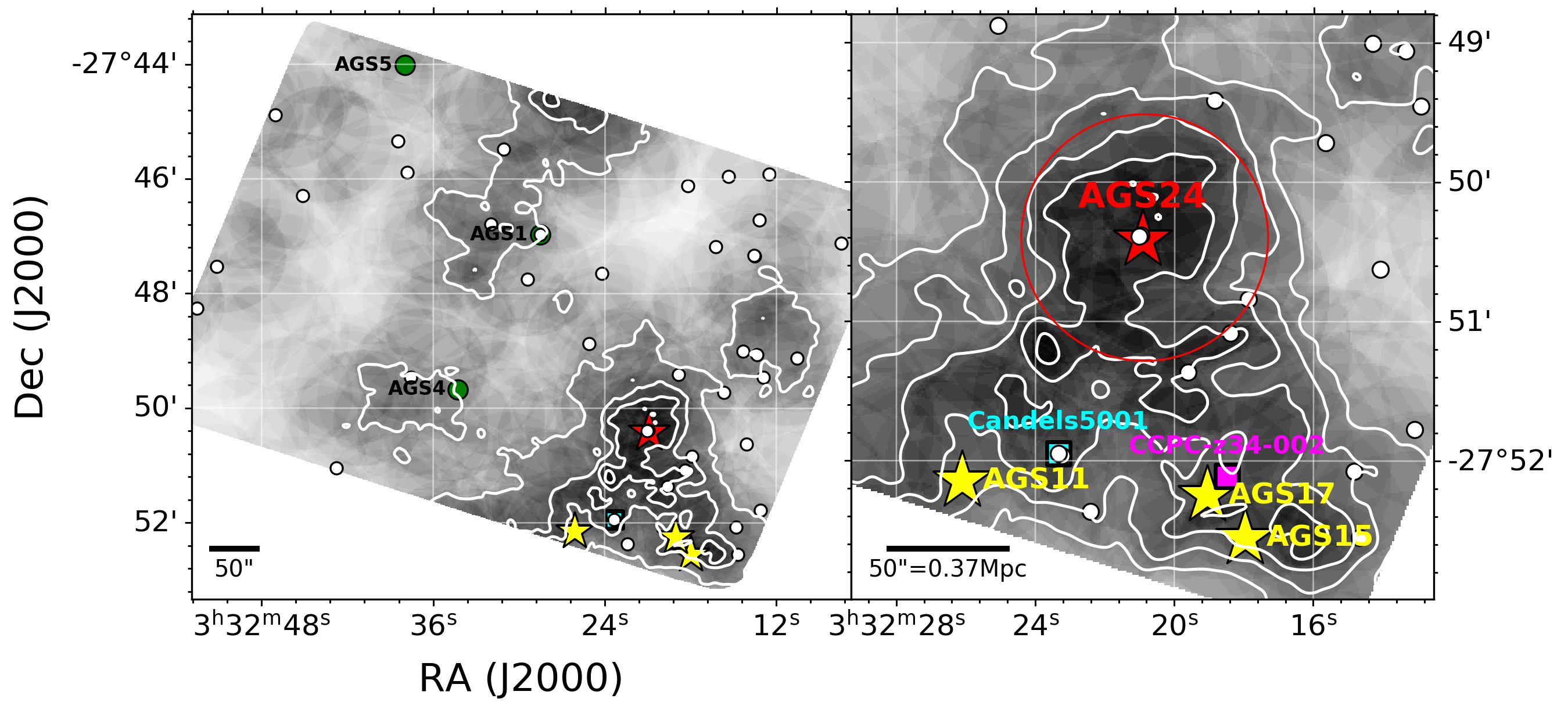}
    \caption{Projected  density of 364 galaxies (40 with $z_{spec}$) at 3.42\,$\leq$\,$z$\,$\leq$\,3.57. The contours show 1,2,3,4$\sigma$ level of the overdensity.  \textbf{\textit{Left}} panel covers the field observed in GOODS-{\it South} ALMA. \textbf{\textit{Right}} panel shows the  zoom-in of the extended structure in the bottom right corner of GOODS-ALMA field that encompasses 126(18) galaxies over 15 arcmin$^2$. The red star indicates the position of AGS24. The red circle indicates a 50\arcsec radius region (2 arcmin$^2$) including 26(3) galaxies. The yellow stars refer to the HST-galaxies AGS 11, 15 and 17 located also in this larger structure. The green circles represent the AGS 1 and 5, which are also in this redshift range but not optically dark. The white circles show the 40 galaxies with spectroscopic redshifts. We note that the two white circles falling on the red star  of AGS24 are two galaxies dissociated from AGS24.  The magenta square denotes the center of the proto-cluster CCPC-z34-002 identified by \citet{Franck16}. The cyan square denotes Candels5001, which used to be the most massive galaxy at 3\,<$\,z$\,<\,4 in this field \citep{Ginolfi17}. The 50\arcsec scale in the figure is equivalent to a proper distance of 0.37 Mpc at $z$\,=\,3.5.}
    \label{fig:density_map}
\end{figure*}

\subsubsection{Projected  density map over the GOODS-ALMA field}
\label{sec:densitymap}

We construct the projected galaxy density in Fig.~\ref{fig:density_map}. Similar to the method used in \citet{Forrest17}, we create a grid with a cell size of 0\farcs9\,$\times$\,0\farcs9 in the GOODS-ALMA field  and calculate the overdensity ($\delta$\,=\,$\rho$\,/\,$\Bar{\rho}$) for each cell.  364 galaxies (40 with $z_{spec}$) at 3.42\,$\leq$\,$z$\,$\leq$\,3.57 are used to produce the map. The galaxies with spectroscopic redshifts are marked as white dots in  Fig.~\ref{fig:density_map}.
An overdense region comes out in the southeast region as indicated by the zoom-in box in Fig. \ref{fig:density_map}. 
The  4$\sigma$ peak of this overdensity happens to be very close to AGS24, the galaxy with the strongest overdensity ($\delta_{\rm AGS24}$\,=\,2.06) this is consistent with the fact that it is in a highly clustered environment.
Out of the 364/40 (redshift/$z_{spec}$) galaxies in the 69 arcmin$^{2}$ field, 126/18 reside in this  15 arcmin$^{2}$ box, and 26/3 galaxies in this redshift slice fall within 50\arcsec\ (2 arcmin$^2$) of AGS24.

We notice that the overdensity shown in the projected  density map  encompasses four out of the six HST-dark galaxies detected by ALMA (AGS 11, 15, 17, 24, stars in Fig.~\ref{fig:density_map}).  The overlap of an overdensity of ALMA detected optically dark galaxies and an overdensity of optically detected galaxies shows that these ALMA detected galaxies are good tracers of structures in the early Universe.  

This overdensity has been classified as a proto-cluster in previous publications. \citet{Franck16} identified it in a list of proto-clusters at 2.74\,<\,$z$\,<\,3.71 built out of $\sim$\,14000 spectroscopic redshifts  (with 604 of them finally used). They defined proto-clusters as galaxy densities greater than the field density by a factor seven within a radius of 2$\arcmin$ and a redshift range of $\delta z$\,<\,0.03. One of their proto-clusters, CCPC-z34-002, has a redshift of $z$\,=\,3.476, hence consistent with the redshift peak of $z$\,=\,3.472 studied here.  We note that the CCPC-z34-002 density peak, roughly estimated from the distribution of only 23 galaxies within a radius of 20\,cMpc, does not correspond to the location of AGS24 but falls only 10$\arcsec$ from AGS17, and the typical diameter of the proto-clusters studied by  \citet{Franck16} of 4\arcmin\ matches the zoom shown in Fig.~\ref{fig:density_map}\textit{-Right}. In the following, we will consider the galaxies at the redshift peak that fall in this area  as a member of a candidate proto-cluster.  This structure encompasses four out of six, hence 67\,\%, of the optically dark galaxies identified in the GOODS-ALMA field. 

This overdensity was also identified by  \citet{Forrest17}  who used a population of extreme emission line galaxies, with low stellar masses and by the VANDELS survey with a three-dimensional algorithm \citep{Guaita2020}.  Moreover, only nine out of 131 the Ly$\alpha$-emitting galaxies selected from the VANDELS survey reside in their 13 detected overdensities at $z$\,>\,2, indicating that Ly$\alpha$-emitting galaxies are not ideal tracers of overdensities.
Finally, we note that  \citet{Ginolfi17} observed the molecular gas content of the galaxy CANDELS-5001 (blue square in Fig.~\ref{fig:density_map}, $M_{\star}$\,=\,10$^{10.27^{+0.38}_{-0.11}}$\,M$_{\odot}$, $z_{spec}$\,=\,3.473) and found that this galaxy was surrounded by an extended emission of molecular gas spanning 40\,kpc traced by the CO(4-3) transition. Furthermore, they detect nine additional CO systems within a radius of 250\,kpc from the massive galaxy and mostly distributed in the same direction as the CO elongated structure found in the central 40\,kpc, which they interpret as evidence for large-scale gas accretion on the galaxy. This large-scale accretion might be enhanced by the larger scale dark matter halo of the structure (see \citealt{Rosdahl2012}).

We calculate the  distance between AGS1, 4, 5 and AGS24 to be 40.5\,cMpc, 70.8\,cMpc, and 25.5\,cMpc, respectively. Despite their larger distance, AGS1 and 5 might be involved in a wider structure than the proto-cluster that includes AGS11, 15, 17, 24 and potentially also AGS4. This wider structure may be compared to the superclusters at high redshift discussed in \citet{Cucciati2018} and \citet{Toshikawa2020}.

\begin{figure}[htbp]
	\includegraphics[width=\hsize]{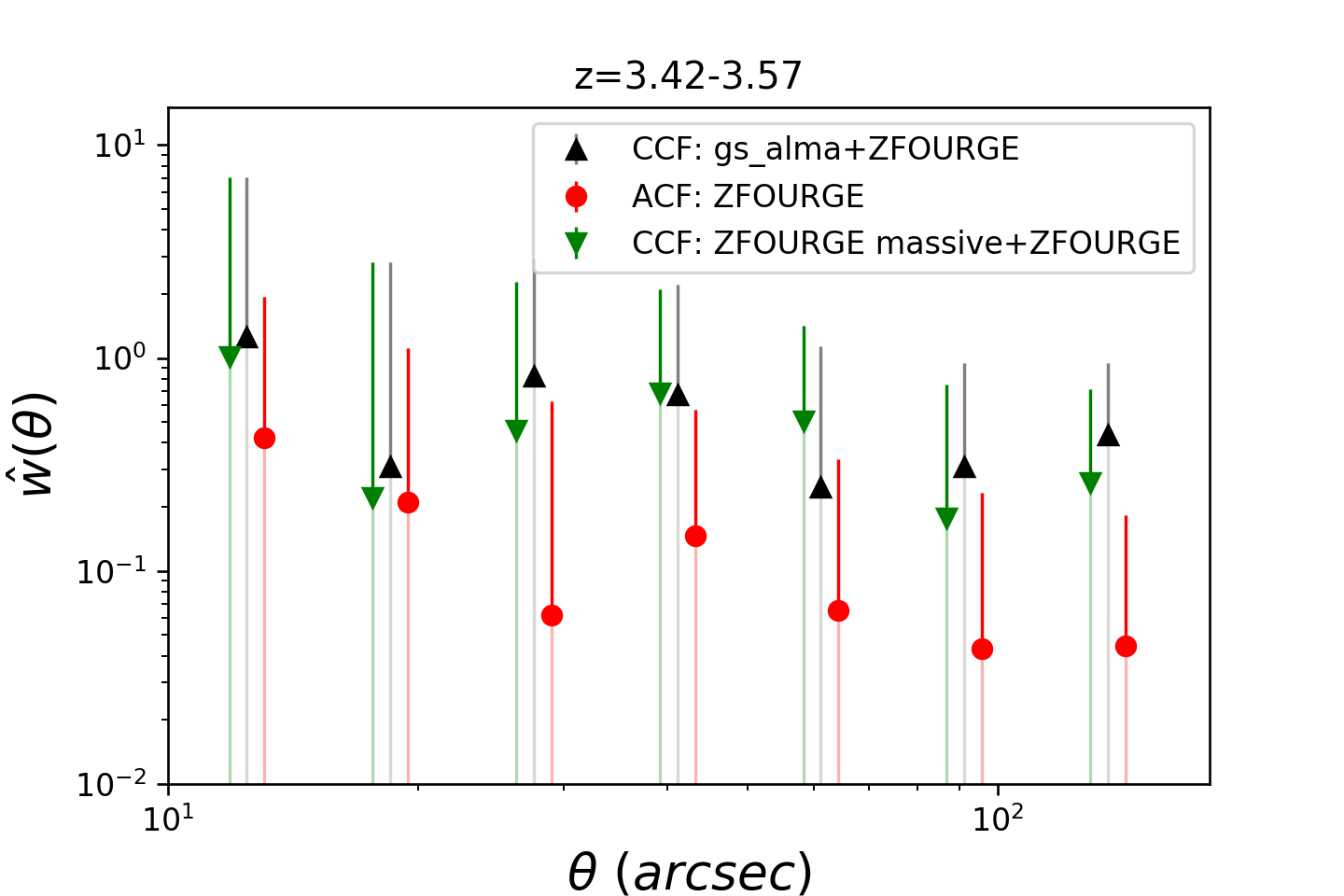}
    \caption{Comparison of the auto-correlation function (ACF) of the  364 galaxies   at  3.42\,$\leq$\,$z$\,$\leq$\,3.57 in the ZFOURGE catalog (red dots), the cross-correlation function (CCF) between these galaxies and the ALMA detections (black triangles), and the CCF between these galaxies and the galaxies that are equivalently massive ({\it M}$_\star$\,>\,10$^{10.5}$\,M$_\odot$}) as the ALMA detections (green triangles).
    \label{fig:CCF}
\end{figure}

\subsubsection{Two point correlation function (2PCF)}

We calculated the auto-correlation function of the 364 galaxies (40 with $z_{spec}$) at 3.42\,$\leq$\,$z$\,$\leq$\,3.57 using a uniformly generated random sample in the GOODS-ALMA field, independent of mass (red dots in Fig. \ref{fig:CCF}), the cross-correlation function between these 364 galaxies and the ALMA detections (black triangles in Fig. \ref{fig:CCF}), and the cross-correlation function between these 364 galaxies and the massive galaxies  that are equivalently massive as the ALMA detections (M$_\star$\,>\,10$^{10.5}$\,M$_\odot$) out of the 364, (green triangles in Fig. \ref{fig:CCF}). A \citet{Landy1993} estimator is adopted. Error bars are estimated from 1000 bootstrap samples. Despite of the large error bars,  the two cross-correlations both show systematically higher excess possibility compared to the auto-correlation of  the 364 galaxies in this redshift range. This indicates a clustering feature of the ALMA detected optically dark galaxies, as well as the equivalently massive galaxies.  


\subsection{Dynamical state of the proto-cluster at $z$\,$\sim$\,3.5}
\label{sec:gal}
As discussed in Section~\ref{sec:massivegal}, AGS24 is the most massive galaxy in the field at $z$\,$>$\,3 ({\it M$_{\star}$}\,=\,10$^{11.32^{+0.02}_{-0.19}}$\,M$_{\odot}$) and in particular at $z$\,$\sim$\,3.5, making it a good candidate for the future BCG of the overdensity of galaxies if it becomes a galaxy cluster. The fact that it is surrounded by three galaxies with the same redshift of $z$\,$\sim$\,3.5 within 1\farcs5 from AGS24 (11.3 kpc at $z$\,$\sim$\,3.5 ) strengthens this hypothesis (Fig. \ref{fig:img_AGS24} and Table \ref{tab:neighbors}). 

To estimate the  dark matter mass of this structure, we first calculated the total stellar mass of all the galaxies within 50\,\arcsec of AGS24 and then determined an upper limit  of $M_h$=10$^{15.0}$\,M$_{\odot}$ to the  dark matter mass based on  the relation in \citet{Behroozi13}, assuming that this structure is already virialized. We also determined a lower limit of  $M_h$=10$^{13.3}$\,M$_{\odot}$  by summing up the individual halo masses of the each galaxy within 50\,\arcsec of  AGS24.

\begin{figure}[htbp]
	\centering
	\includegraphics[width=3.0in]{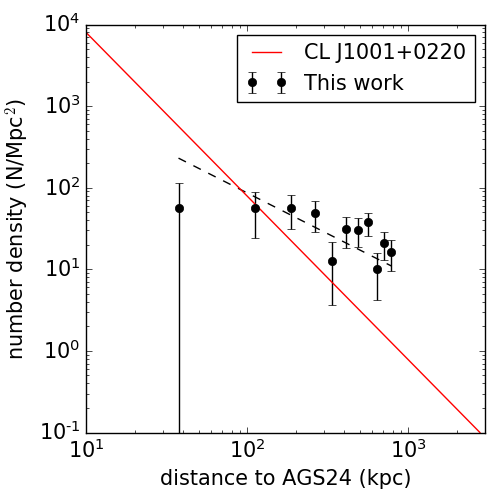}
    \caption{Projected numbers of galaxies in the overdensity as a function of their distance to AGS24 (black circles) and the black dashed line is the linear fitting to the profile. 
    The red line shows the best-fit projected NFW profile  \citep{Navarro96}  of a galaxy cluster at $z$\,=\,2.506, CL J1001+0220  \citep{Wang16}.}
    \label{fig:profile}
\end{figure}

We note that the  projected number density profile of the galaxies in the structure do not follow the slope expected for the NFW profile \citep{Navarro96} of a gravitationally bound cluster, but instead exhibit a shallow slope (Fig.~\ref{fig:profile}). We show for comparison the projected number density profile of the most distant galaxy cluster to our knowledge, CL\,J1001+0220 at $z$\,=\,2.506 \citep{Wang16}. 

However, the structure shown in the density map (Fig.~\ref{fig:density_map}), as well as the fact that AGS24 is the most massive galaxy that falls at the density peak and has a reduced SFR compared to MS galaxies, suggests that this structure is in the process of viralization and that such an environment is influencing the evolution of AGS24, diminishing its star formation activity. Galaxies like AGS24 have been missed by previous studies given their dusty nature, despite being the  most massive candidate galaxies that can be used as a beacon to trace the barycenter of the proto-clusters in the process of virialization.

\section{Conclusions}
We have  analyzed the properties of an ensemble of six optically dark galaxies, AGS4, 11, 15, 17, 24 and 25 identified in  the GOODS-ALMA survey. They do not have optical counterparts in the deepest $H$-band based catalog down to $H$\,=\,28.16\,AB.  
Out of the six sources initially classified as optically dark, a careful analysis revealed that two were associated with an $H$-band counterpart that was not identified in the $H$-band catalog missed due to blending. AGS4 and AGS15, that were identified as optically dark in  \citetalias{Franco18}, are  extremely close to a bright optical source at 0\farcs50 and 0\farcs27,  respectively, and as such they were wrongly considered as parts of the bright neighbor. However, we showed that their SED can be used to perform a spectral de-confusion of the sources at very different redshifts. We note that the $H$-band magnitude of these galaxies is extremely faint since it is 25.2\,AB and 27.1\,AB for AGS4 and AGS15, respectively. The second one is below the detection limit of most $H$-band images. The definition of HST-dark or optically dark is obviously a function of the depth of the HST images.
All of six sources except AGS24 do exhibit a counterpart in the $K_s$-band image but they were mistakenly associated with a bright neighbor, hence none was listed in the ZFOURGE catalog.

We performed a spectroscopic scan follow-up of five of them with ALMA (AGS4, 11, 15, 17 and 24). All five exhibit a clear 2\,mm counterpart in the continuum with $S/N$\,=\,14.7, 16.8, 15.3, 15.9, and 8.7, respectively, which reinforces their robustness. We detected one emission line for AGS4  ($\nu_{obs}$\,=\,151.44\,GHz with a $S/N$\,=\,8.58) and for AGS17 ($\nu_{obs}$\,=\,154.78\,GHz with a $S/N$\,=\,10.23). Taking into consideration the PDF of the photometric redshifts derived from the optical to near-infrared SEDs and the far-infrared SEDs as well as the infrared luminosities, we conclude that the spectroscopic redshifts of the two galaxies match the CO(6-5) transition and are $z^{\rm AGS4}_{spec}$\,=\,3.556 and $z^{\rm AGS17}_{spec}$\,=\,3.467. Existing spectroscopic confirmation of optically dark sources have been so far limited to a few sources, such as, $z$\,=\,5.183: \citealt{Walter12}; $z$\,=\,3.717: \citealt{Schreiber18b}; $z$\,=\,3.097 \& 5.113: \citealt{Wang19}.

Close to 70\,\% (4/6) of the optically dark galaxies reside in an overdensity of galaxies at $z$\,$\sim$\,3.5 (AGS11, 15, 17 and 24, Fig.~\ref{fig:density_map}). In \citet{Wang19},  the HST-dark galaxies  exhibit a highly heterogeneous spatial distribution illustrated by their strong clustering measured by the cross-correlation functions. This is consistent with the present finding of a strong association of most of the blindly detected optically dark galaxies in GOODS-ALMA of four out of six galaxies located in the same proto-cluster. 
In addition, the redshifts of these galaxies are all  consistent with this overdensity (Table.~\ref{tab:hstdark}), indicating that these galaxies are tracing a galaxy cluster in formation. 

We also notice that AGS24 is not only in the barycenter of the structure at $z$\,$\sim$\,3.5 but also the most massive galaxy in this structure ({\it M$_{\star}$}\,=\,10$^{11.32^{+0.02}_{-0.19}}$\,M$_{\odot}$), suggesting that AGS24 is a candidate BCG in formation. In fact, AGS24 is the most massive galaxy without an AGN at $z$\,>\,3  in the GOODS-ALMA field.  Compared to the other five optically dark galaxies  with high starburstness, AGS24 is the only MS galaxy (Table~\ref{tab:hstdark}) indicating that it is influenced by the environment it resides in.  The fact that AGS24 is a is a candidate BCG also indicates that the proto-cluster is in the process of virialization.


\begin{acknowledgements}
We thank the anonymous referee for constructive comments that helped improve the overall quality and consistency of the paper. 
We thank  Wiphu Rujopakarn, Kristina Nyland and  Preshanth Jagannathan  for sharing the HUDF 3 GHz image.
We thank  Francesco Valentino for providing information on the CO(7-6) and [CI](2-1) emission of existing observations.
We thank Ivan Delvecchio for helpful discussions on radio galaxies. 
\newline
L.Z.  and Y.S. acknowledge the support from the National Key R\&D Program of China (No. 2017YFA0404502, No. 2018YFA04027040404502) and the National Natural Science Foundation of China (NSFC grants 11825302, 11733002 and 11773013). L.Z.  also acknowledges  China Scholarship Council (CSC).
This work was supported by the Programme National Cosmology et Galaxies (PNCG) of CNRS/INSU with INP and IN2P3, co-funded by CEA and CNES.
M.F. acknowledges support from the UK Science and Technology Facilities Council (STFC) (grant number ST/R000905/1)
R.D. gratefully acknowledges support from the Chilean Centro de Excelencia en Astrof\'isica y Tecnolog\'ias Afines (CATA) BASAL grant AFB-170002.
H.I. acknowledges support from JSPS KAKENHI Grant Number JP19K23462.
M.P. is supported by the ERC-StG 'ClustersXCosmo', grant agreement 71676
\newline
VANDELS survey is based on data products created from observations collected at the European Organisation for Astronomical Research in the Southern Hemisphere under ESO programme 194.A-2003(E-T).
\newline
This paper makes use of the following ALMA data: ADS/JAO.ALMA  \#2015.1.00543.S and \#2018.1.01079.S. ALMA is a partnership of ESO (representing its member states), NSF (USA) and NINS (Japan), together with NRC (Canada), MOST and ASIAA (Taiwan), and KASI (Republic of Korea), in cooperation with the Republic of Chile. The Joint ALMA Observatory is operated by ESO, AUI/NRAO and NAOJ.
\newline
This paper employed  \texttt{Astropy}\footnote{\url{http://www.astropy.org}}, a community-developed core Python package for Astronomy \citep{astropy2013, astropy2018}; \texttt{APLpy}, an open-source plotting package for Python \citep{aplpy2012, aplpy2019}; \texttt{Matplotlib} \citep{Hunter2007}; \texttt{Numpy} \citep{oliphant2006guide};  \texttt{SciPy} \citep{2020SciPy-NMeth}; \texttt{CASA} \citep{McMullin2007}; \texttt{GILDAS}\footnote{\url{http://www.iram.fr/IRAMFR/GILDAS}} \citep{Gildas}. 
\end{acknowledgements}

\bibliographystyle{aa}
\bibliography{bibtex}

\end{document}